\RequirePackage{fix-cm}
\documentclass[twocolumn]{svjour3}          % twocolumn
\smartqed  % flush right qed marks, e.g. at end of proof
\usepackage{amsmath,amssymb,nicefrac,setspace}
\usepackage{amssymb,cite,sidecap,isomath}
\usepackage{graphics}
\usepackage[hidelinks]{hyperref}
\usepackage{tikz,enumerate}
\usepackage{subcaption}

\usepackage{graphicx}
\begin{document}

\title{Unstable Limit Cycles Estimation from Small Perturbations: A Model-Free Approach%\thanks{Grants or other notes about the article that should go on the front page should be placed here. General acknowledgments should be placed at the end of the article.}
}


\author{Giuseppe Habib$^*$}


\institute{G. Habib (Corresponding author) \at
              a) Department of Applied Mechanics, Faculty of Mechanical Engineering, Budapest University of Technology and Economics, M\H{u}egyetem rkp. 3., H-1111 Budapest, Hungary. \\
b) MTA-BME Lend\"ulet ``Momentum" Global Dynamics Research Group, Budapest University of Technology and Economics, M\H{u}egyetem rkp. 3., H-1111 Budapest, Hungary.
              Tel.: +36-1-463-1369\\
              \email{habib@mm.bme.hu}  \\
              Orcid: 0000-0003-3323-6901}

\date{Received: date / Accepted: date}

\maketitle

\begin{abstract}

Although stable solutions of dynamical systems are typically considered more important than unstable ones, unstable solutions have a critical role in the dynamical integrity of stable solutions. In fact, usually, basins of attraction boundaries are composed of unstable solutions and their stable manifolds.
This study proposes a method for estimating unstable limit cycles surrounding stable equilibrium points. The method exploits the shape of the decrement of trajectories converging towards the equilibrium. For the method, trajectories obtained from small perturbations from the equilibrium state are sufficient to estimate the unstable limit cycle roughly.
No mathematical model of the system dynamics is needed for the computation, which requires only a single trajectory in the phase space.
As such, the method is computationally very rapid and potentially implementable in real structures.

\end{abstract}

%%%%%%%%%%%%%%%%%%%%%%%%%%%%%%%%%%%%%%%%%%%%%%%%%%%%%%%%%%%%%%%%%%%%%%%%%%%%%%%%%%%%%%%%%%%%%%%%%%%%%%%%%%%%%%%%%%%%%%%%%%%

\section{Introduction}

[For a detailed introduction, including a thorough literature review, please see the original publication (currently still unavailable)].

This study introduces a method for estimating unstable limit cycles (ULCs) of dynamical systems. The method leverages the so-called critical slowing down \cite{strogatz2018nonlinear} to estimate the extent of a ULC surrounding a stable equilibrium point. The estimation is performed by interpreting how a trajectory converges toward the equilibrium after the system undergoes a small perturbation.

The main working principle of the method is discussed in Sect.~\ref{sect_idea}. In Sect.~\ref{sect_method}, the details of the method are explained. The method is then tested on three systems presenting very different challenges. Namely, the mass-on-moving-belt system, which presents non-\-smooth\-ness (Sect.~\ref{sect_MOB}), a time-delayed mathematical model for turning machining dynamics (Sect.~\ref{sect_delay}), and a pitch-and-plunge wing profile undergoing flutter instability, whose specific challenge is the presence of more than one degree-of-freedom (DoF) (Sect.~\ref{sect_PP}).
Concluding remarks are provided in Sect.~\ref{sect_conclusions}.

\section{Fundamental idea}\label{sect_idea}

Let us first consider a slightly damped, unforced linear oscillator equivalent to the classical spring-mass-damper system.
For any given perturbation of the system from the trivial solution, it oscillates with vibrations of decreasing amplitude. The amplitude decrement is well described by the logarithmic decrement $\Lambda$, which can be defined as \begin{equation}
\Lambda=\frac{A_i}{A_{i+1}},
\end{equation}
where $A_i$ is the oscillation amplitude of the $i^{\text{th}}$ peak.
For a linear system, $\Lambda$ is constant, and its value is related to the damping ratio of the system.

\begin{figure*}[t]
\begin{center}
\setlength{\unitlength}{\textwidth}
\begin{subfigure}[b]{0.325\textwidth}
    \includegraphics[width=\textwidth]{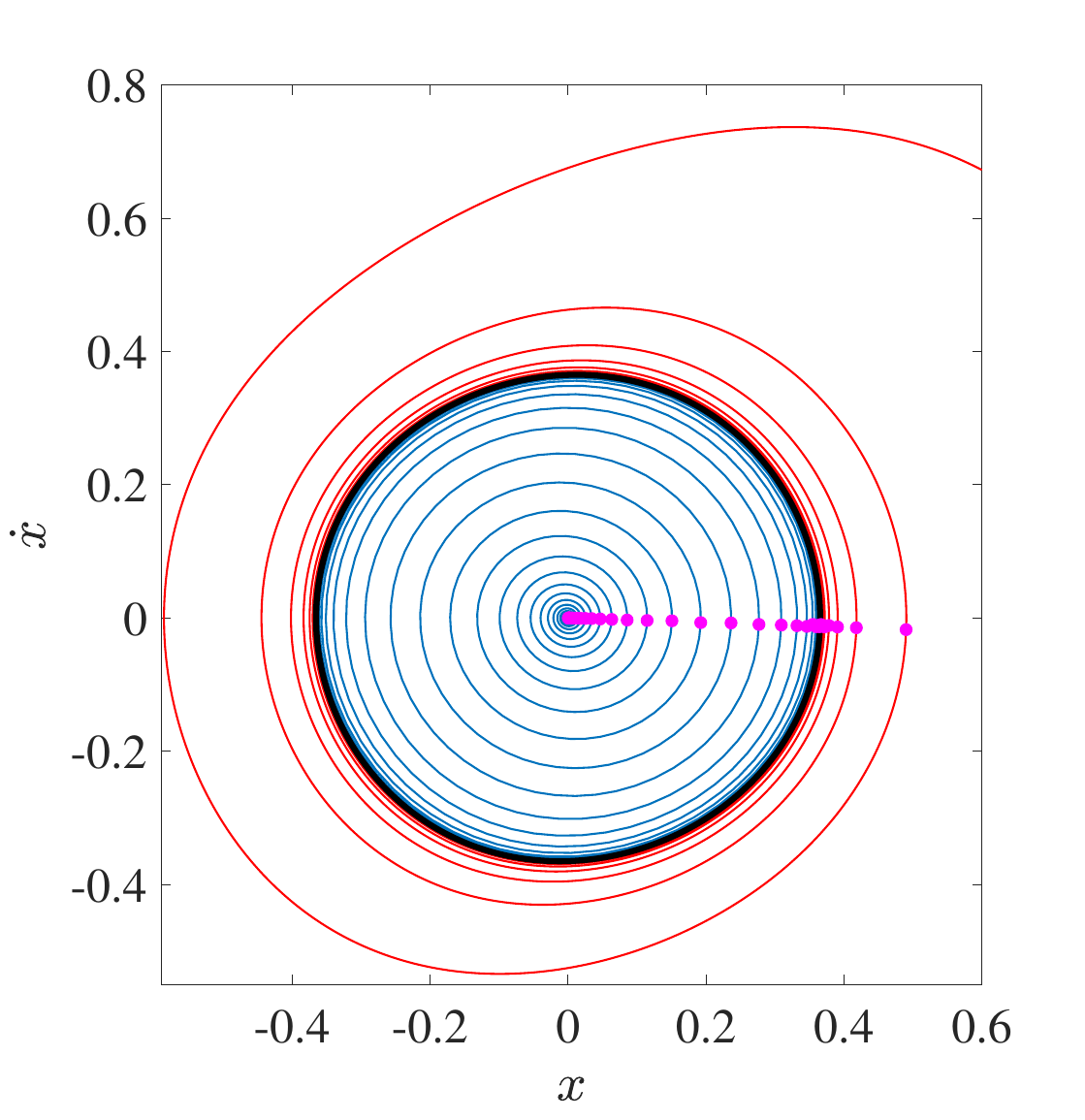}
    \caption{}
    \label{fig_PP_example}
  \end{subfigure}
\begin{subfigure}[b]{0.325\textwidth}
    \includegraphics[width=\textwidth]{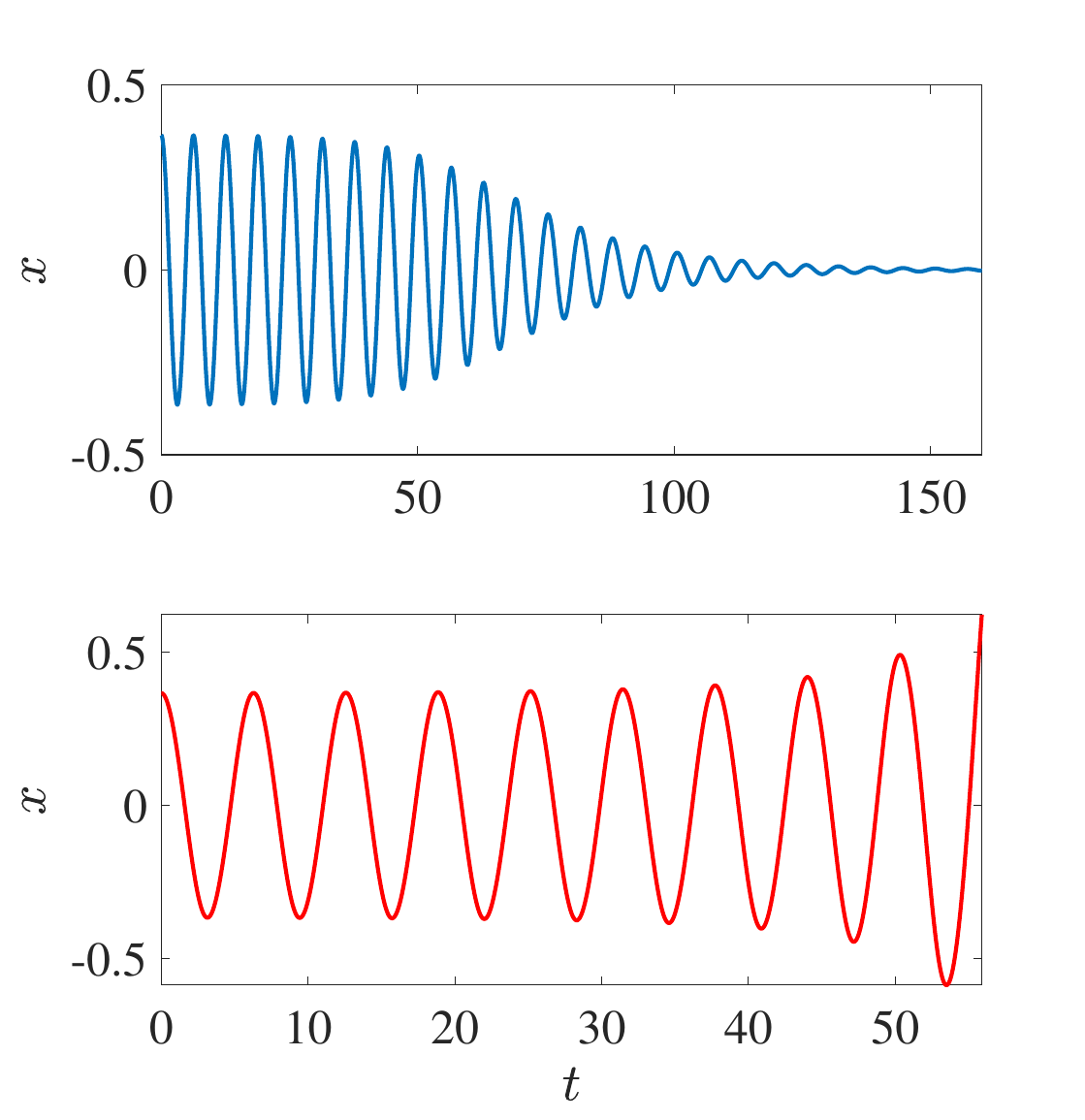}
    \caption{}
    \label{fig_time_series_example}
  \end{subfigure}
  \begin{subfigure}[b]{0.325\textwidth}
    \includegraphics[width=\textwidth]{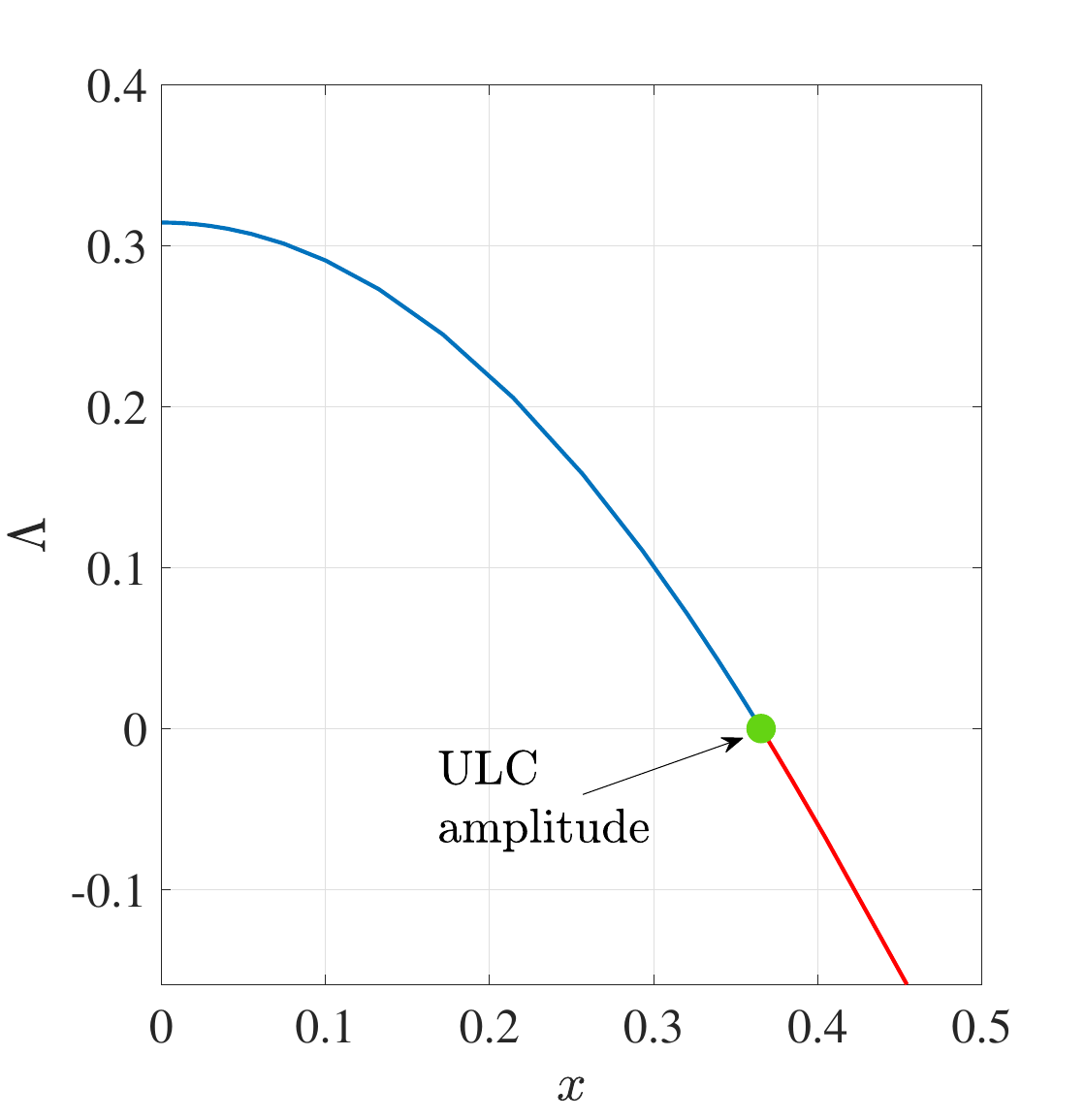}
    \caption{}
    \label{fig_LD_example}
  \end{subfigure}
\end{center}
\caption{\label{fig_example}(a) Phase portrait of a system having a stable trivial solution and a ULC (black circle); blue curve: trajectory converging to the equilibrium; red curve: trajectory diverging from it; magenta dots: intersections of trajectories with the positive zero semi-axis. (b) Time series either converging towards a stable equilibrium (top), or diverging from it (bottom). (c) Logarithmic decrement computed for each peak of the trajectories; blue curve: converging trajectory, red curve: diverging trajectory; green dot: amplitude of the ULC.}
\end{figure*}

Now, let us consider a single-DoF smooth nonlinear system having a stable equilibrium surrounded by a ULC, as illustrated in Fig.~\ref{fig_PP_example}.
The limit cycle is a solution of the system. Accordingly, despite being unstable, a trajectory passing close to it in the phase space moves away relatively slowly.
This phenomenon is often referred to as critical slowing down \cite{strogatz2018nonlinear}, which means that the dynamics of a system tends to critically slow down in the vicinity of steady-state solutions despite their stability properties.
In this sense, the velocity of the system is interpreted as the rapidity to diverge from or to converge toward the solution.

Let us consider a trajectory starting from the vicinity of the unstable limit cycle and converging toward the trivial solution. The relative amplitude decrement, quantified by the logarithmic decrement $\Lambda$, will no longer be constant, as for the linear system, but will increase from a small value near the ULC, up to a larger value while approaching the equilibrium.
If the trajectory is initiated from a point infinitesimally close to the unstable solution, the logarithmic decrement will start from almost zero. If the initial condition is inside the ULC, oscillation amplitude will decrease, and $\Lambda$ will increase.
Conversely, if the initial condition lies outside the ULC, the amplitude will increase, and the logarithmic decrement will be negative and decrease.
Graphically, this corresponds to a curve intersecting zero in correspondence with the amplitude of the ULC.
This phenomenon is illustrated in Fig.~\ref{fig_LD_example}.

Through a perturbation of the system from its equilibrium condition, part of the curve marking the trend of the logarithmic decrement can be calculated.
If the trend of the logarithmic decrement is relatively regular and can be approximated by a polynomial curve, then its intersection with the zero axis can be estimated, as discussed below in Sect.~\ref{sect_method}.
Although the method is more straightforward for a single-DoF system, it can also be applied to larger dynamical systems, as illustrated in Sects.~\ref{sect_multiDoF} and \ref{sect_PP}.
The formalization of the method for estimating ULC is described in Sect.~\ref{sect_method}.

\section{Unstable limit cycle estimation} \label{sect_method}

\subsection{Single-DoF systems}\label{sect_meth_1DoF}
\begin{figure*}[t]
\begin{center}
\setlength{\unitlength}{\textwidth}
\begin{subfigure}[b]{0.325\textwidth}
    \includegraphics[width=\textwidth]{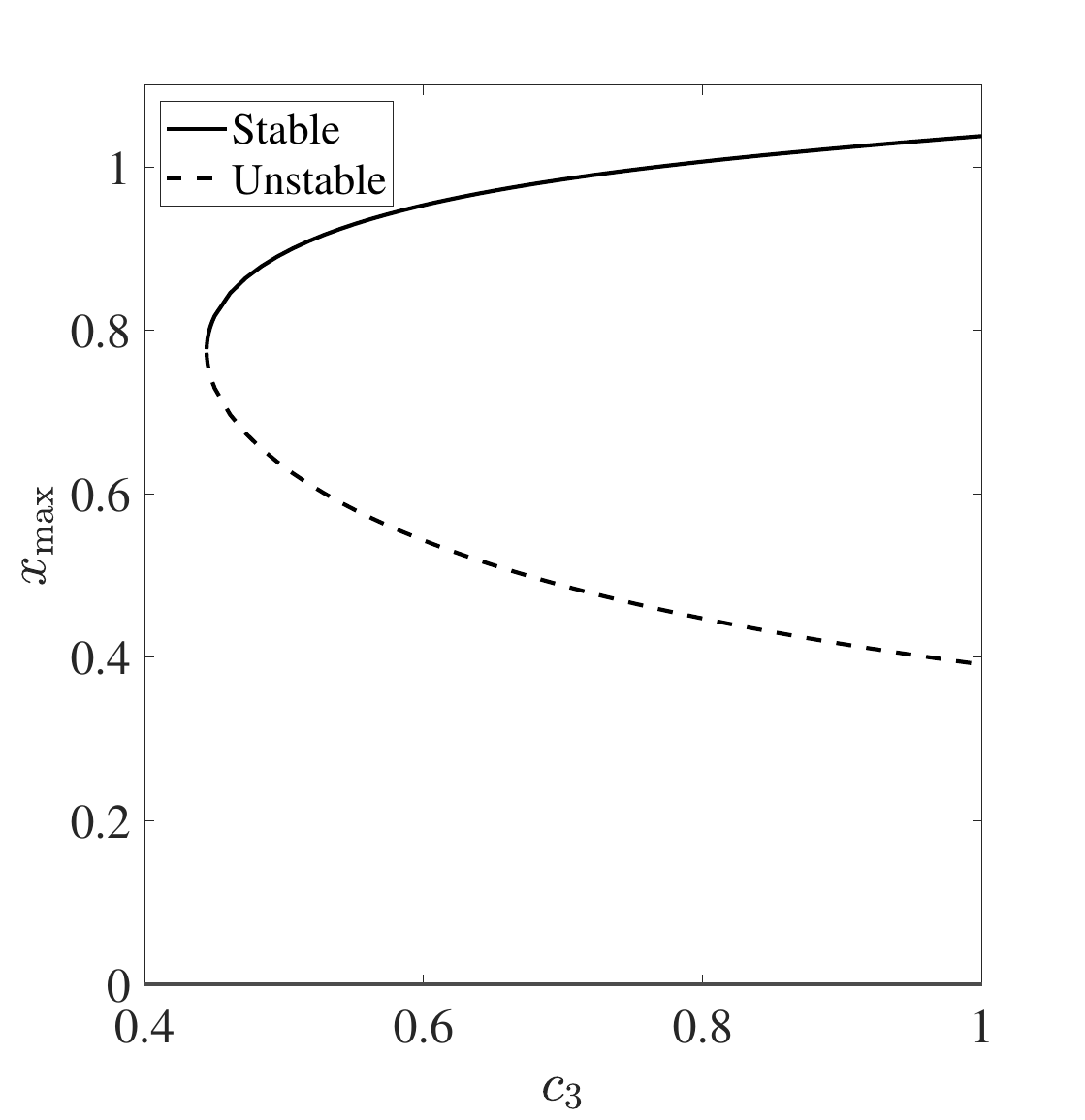}
    \caption{}
    \label{fig_bif_diag_ND}
  \end{subfigure}
\begin{subfigure}[b]{0.325\textwidth}
    \includegraphics[width=\textwidth]{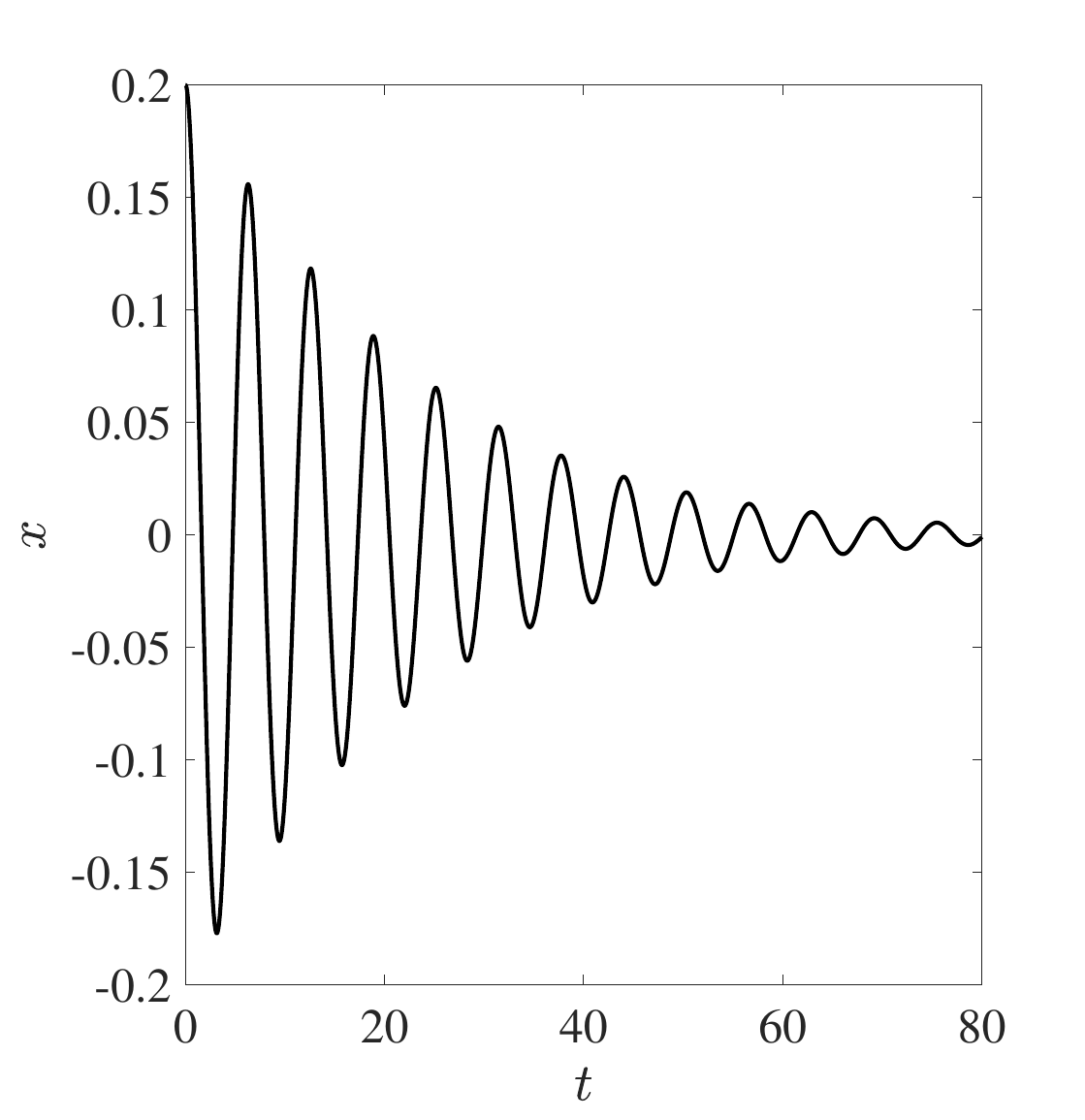}
    \caption{}
    \label{fig_time_series_ND}
  \end{subfigure}
  \begin{subfigure}[b]{0.325\textwidth}
    \includegraphics[width=\textwidth]{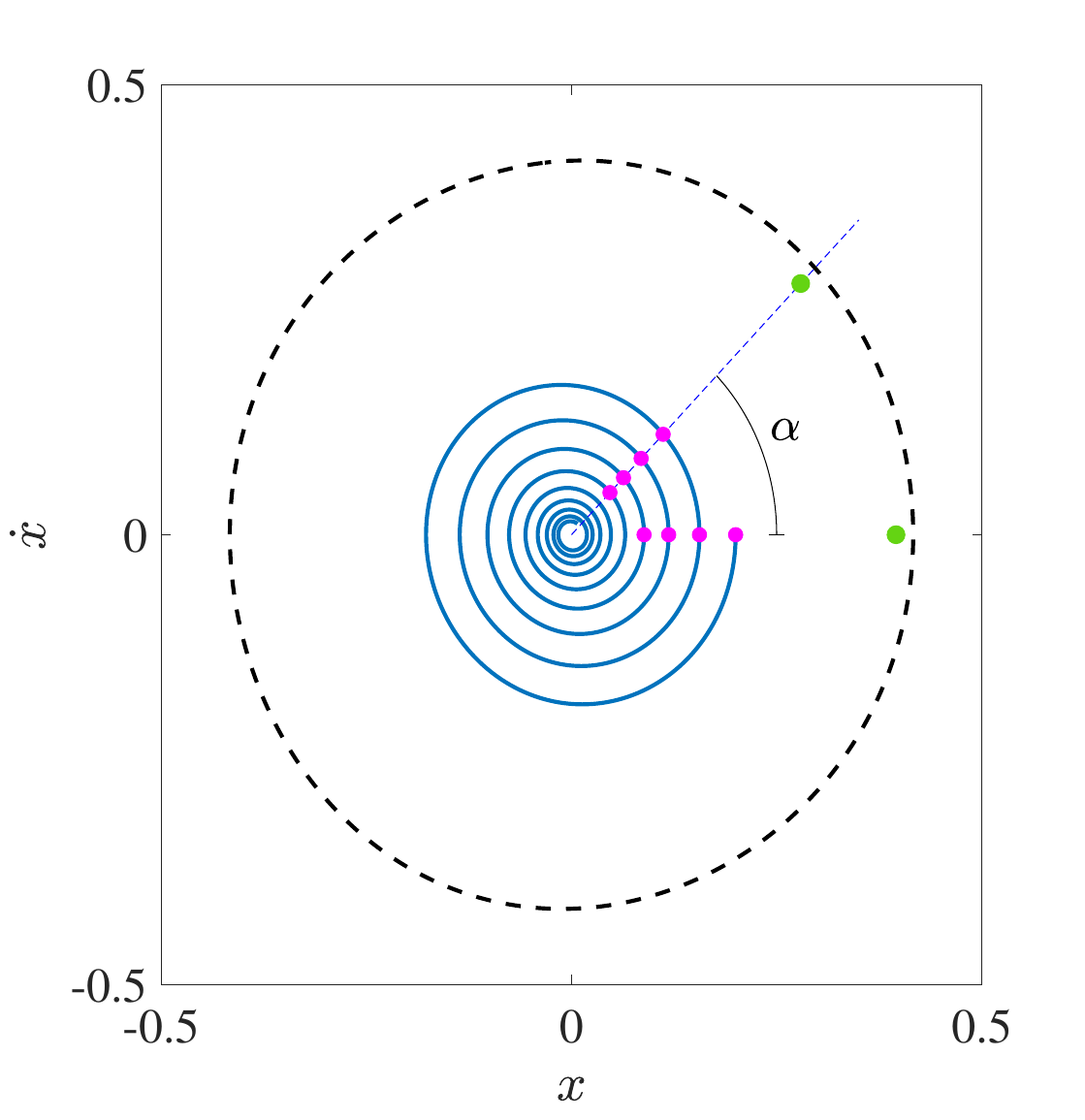}
    \caption{}
    \label{fig_PP_ND}
  \end{subfigure}
  \begin{subfigure}[b]{0.325\textwidth}
    \includegraphics[width=\textwidth]{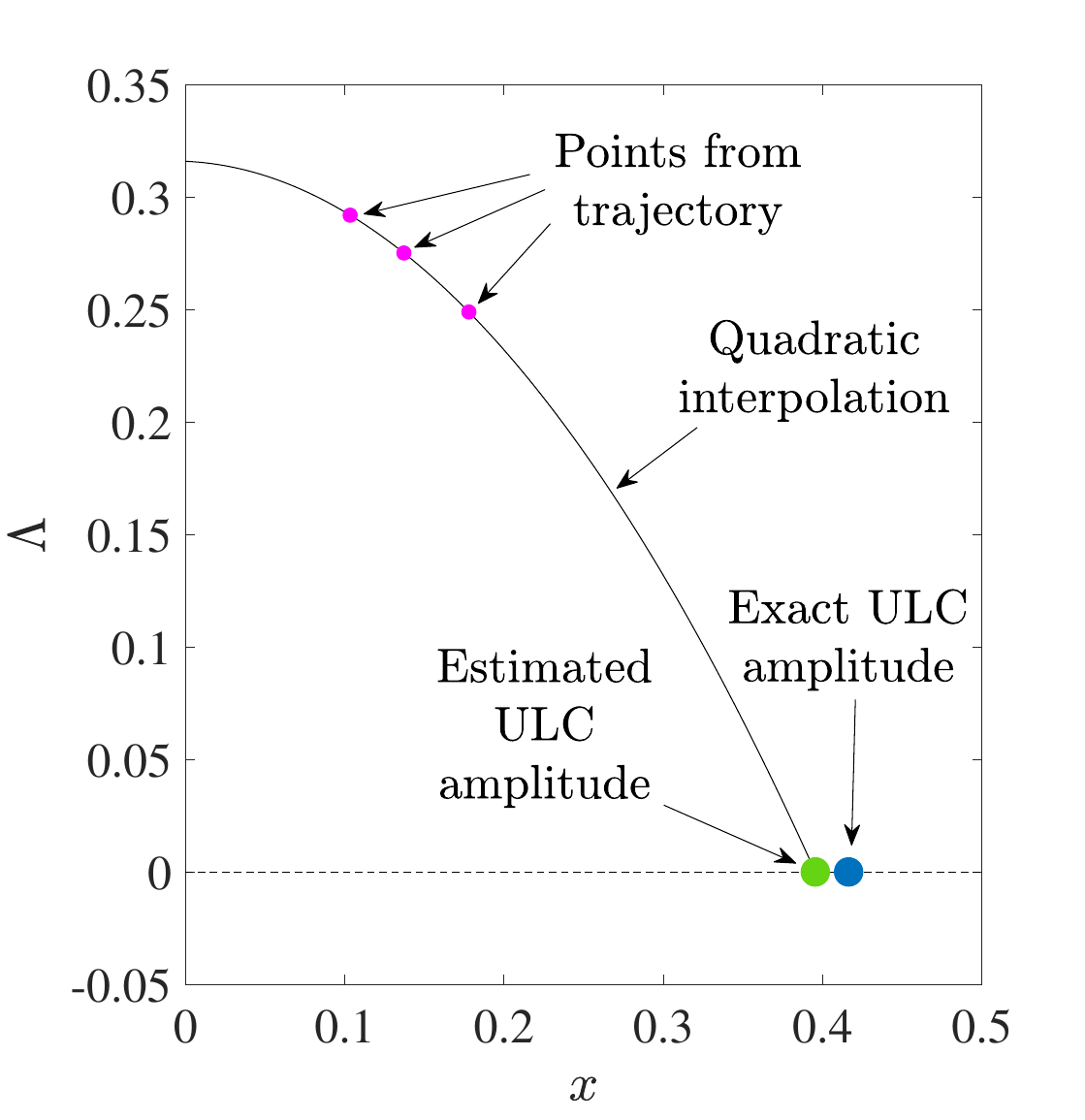}
    \caption{}
    \label{fig_LD_ND}
  \end{subfigure}
  \begin{subfigure}[b]{0.325\textwidth}
    \includegraphics[width=\textwidth]{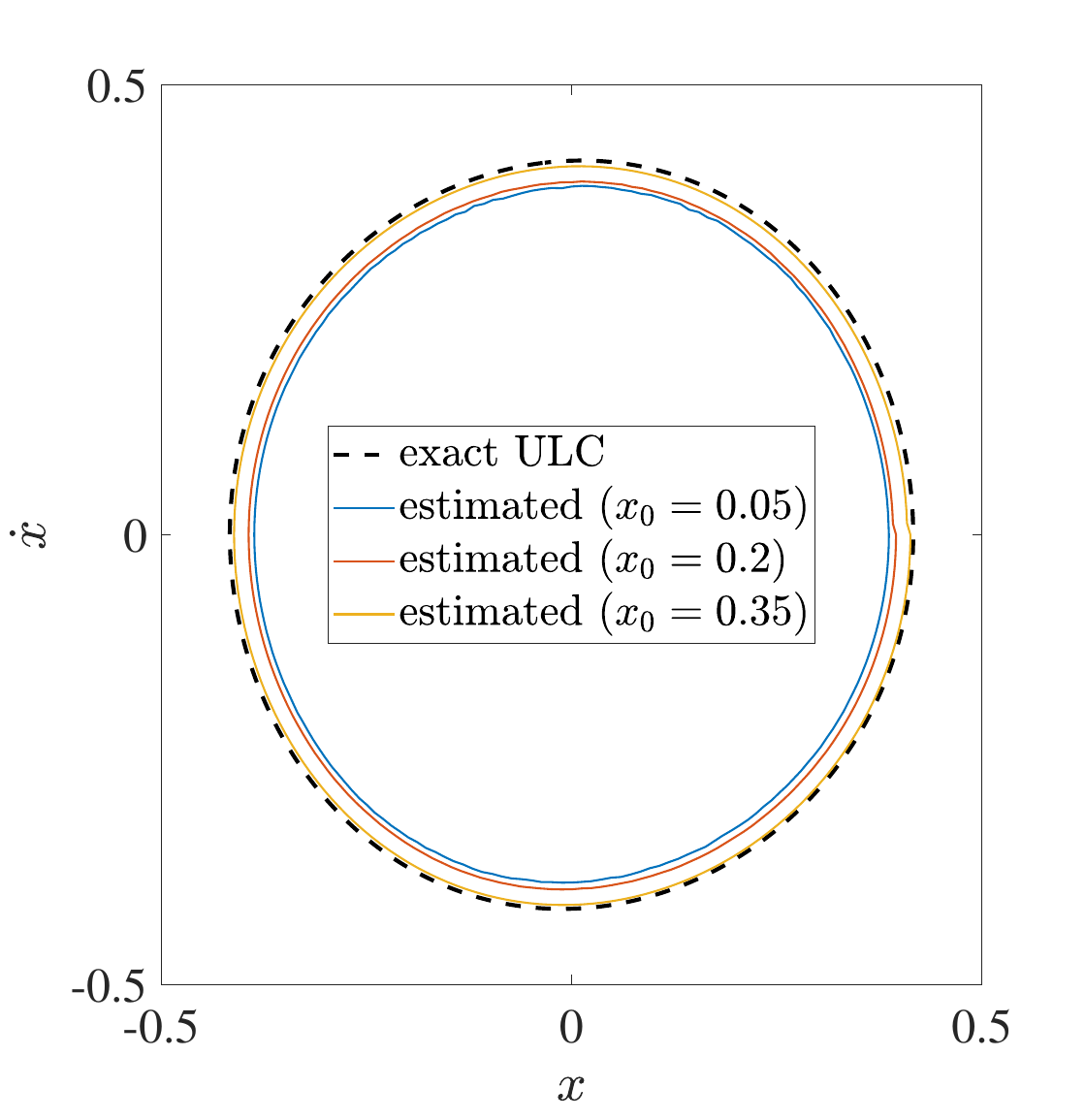}
    \caption{}
    \label{fig_PP_pred_ND}
  \end{subfigure}
  \begin{subfigure}[b]{0.325\textwidth}
    \includegraphics[width=\textwidth]{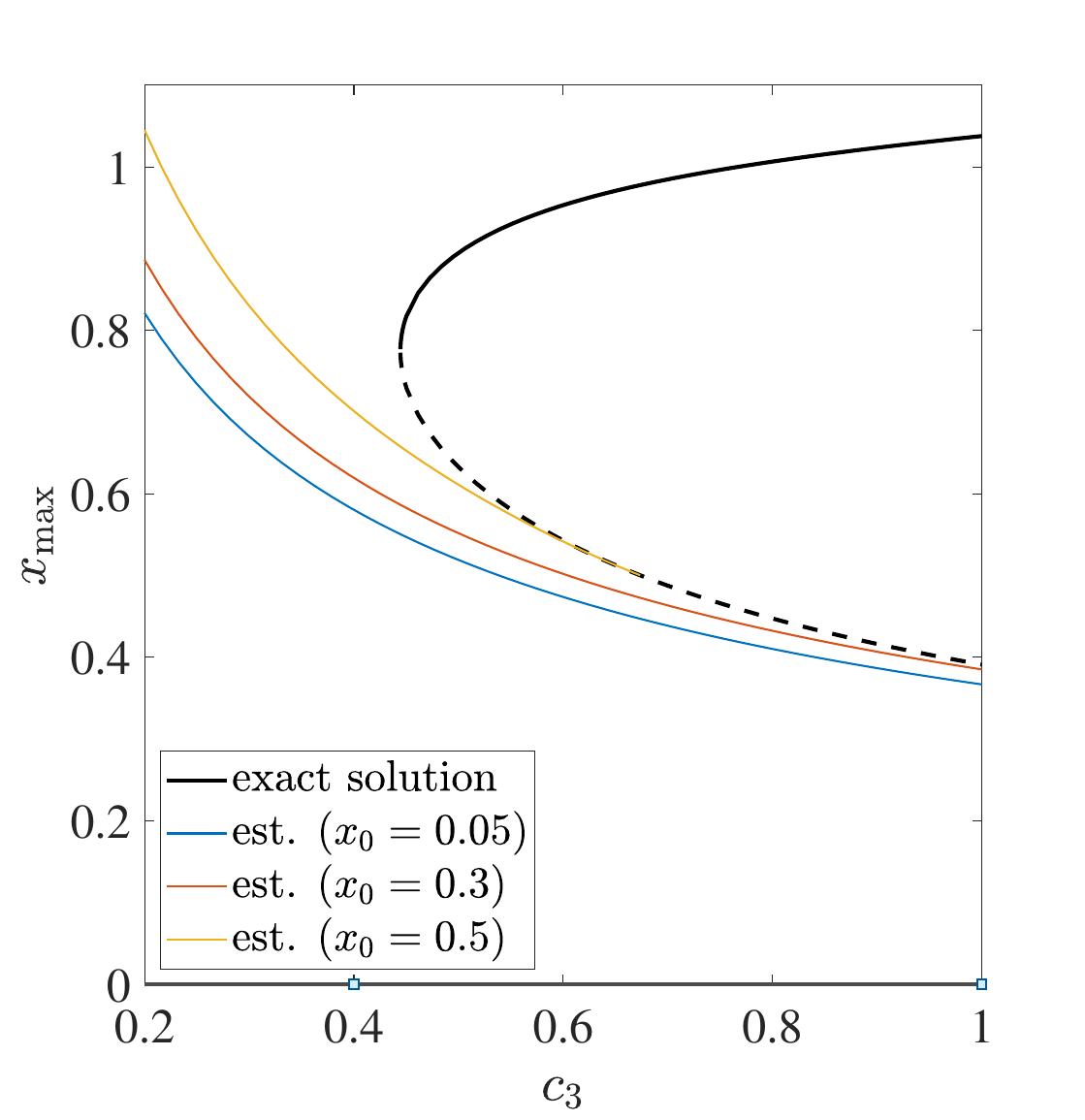}
    \caption{}
    \label{fig_bif_pred_ND}
  \end{subfigure}
\end{center}
\caption{\label{fig_example}(a) Bifurcation diagram for the system in Eq.~(\ref{eq_EoM_ND}), $c_1=0.1$. (b) Time series for the ULC estimation, obtained for $c_3=0.9$. (c) Phase portrait of the system for $c_3=0.9$; blue curve: simulated trajectory; black dashed curve: exact ULC; blue dashed line: Poincar\'e section; magenta dots: points of the trajectory selected for the ULC estimation; green dot: estimated ULC for either $\dot x=0$ and on the Poincar\'e section. (d) Logarithmic decrement point and ULC estimation. (e) Comparison between exact and estimated ULC for different initial conditions. (f) Comparison between the exact and estimated unstable branch of the bifurcation diagram for different initial conditions.}
\end{figure*}

Let us consider a single-DoF system possessing a stable trivial solution and a ULC. For illustrative purposes, we consider a specific system, i.e., an oscillator possessing a nonlinear damping characteristic, whose dynamics is governed by the differential equation \begin{equation}
\ddot x+x+c_1\dot x-c_3\dot x^3\left(1-\dot x^2\right)=0, \label{eq_EoM_ND}
\end{equation}
where $x$ is the state variable, $c_1$ is the linear damping coefficient and $c_3$ is the nonlinear damping coefficient.
This system was thoroughly studied in \cite{habib2018isolated}.
For $c_1>0$, the trivial solution is always stable. However, while for $c_3<c_3^*=40c_1/9$ it is also globally stable, for $c_3>c_3^*$ it coexists with a stable and an unstable limit cycle. The unstable limit cycle marks the boundary between the basins of attraction of the two solutions. $c_3=c_3^*$ corresponds to a fold bifurcation where the stable and unstable solution branches merge. The corresponding bifurcation diagram is illustrated in Fig.~\ref{fig_bif_diag_ND}.
For the analysis, the parameter range of interest is for $c_3>c_3^*$, where multiple solutions coexist. Without loss of generality, in the following analysis, $c_1$ is fixed at 0.1.

A small perturbation is imposed on the system, moving it away from its trivial equilibrium; however, it is kept within the basin of attraction of the trivial solution, i.e., inside the area spanned by the ULC.
Time series and phase portrait of the ensuing dynamics are depicted in Figs.~\ref{fig_time_series_ND} and \ref{fig_PP_ND}.
The peaks of the time series in Fig.~\ref{fig_time_series_ND} correspond to the intersection of the trajectory with the zero axis, marked by magenta dots in Fig.~\ref{fig_PP_ND}.
The first four peaks are considered, and the logarithmic decrement between them is computed, such as: \begin{equation}
\Lambda_i=\frac{A_i}{A_{i+1}}.
\end{equation}
Then, logarithmic decrement values $\Lambda_i$ are associated with the average amplitude of the peaks for which they were computed, obtaining the three couple of points: \begin{equation}
\left(\frac{A_i+A_{i+1}}{2},\Lambda_i\right),
\end{equation}
as depicted in Fig.~\ref{fig_LD_ND} (magenta dots).
Only the first three points encountered are used since they provide the best estimation because they are the closest to the ULC.

After that, the quadratic polynomial passing through the three points is computed (black curve in Fig.~\ref{fig_LD_ND}), and its intersection with the $\Lambda=0$ axis is calculated. The amplitude where the intersection occurs is the estimated amplitude of the ULC (green dot in Fig.~\ref{fig_LD_ND}).

In order to find the whole ULC, and not only its intersection with the $\dot x=0$ axis, the procedure can then be repeated for various Poincar\'e sections of the phase space, each passing through zero and having a different angle $\alpha$ with respect to the $\dot x=0$ axis, as illustrated in Fig.~\ref{fig_PP_ND}.
For each Poincar\'e section, the first four intersections of the trajectory are picked and treated exactly as for the case of the peaks $A_i$, from which the corresponding amplitude of the ULC on the Poincar\'e section can be estimated.
If the angle $\alpha$ spans $2\pi$ radiants, the full limit cycle is estimated.
Very unusual shapes of the ULC, involving multiple crossing of the Poincar\'e section in a single loop, lead to the failure of the procedure. However, such cases are not very common; therefore, this limitation is not critical. Besides, transformations of the system coordinates might mitigate the issue.

Figure \ref{fig_PP_pred_ND} compares estimated ULCs with the exact one for different perturbation amplitudes. We remark that even very small perturbations are able to provide very good estimations of the ULC, as the estimated ULC for $x(0)=0.05$ clearly illustrates.

Repeating the procedure for a range of the bifurcation parameter ($c_3$ in this case), estimating the full bifurcation diagram is possible.
Estimated bifurcation diagrams, obtained from perturbations of various amplitudes, are compared with the exact one in Fig.~\ref{fig_bif_pred_ND}. The estimation was interrupted if the trajectory converged towards the stable limit cycle and not to the equilibrium.
From Fig.~\ref{fig_bif_pred_ND}, the relatively good accuracy of the estimation can be appreciated. However, we note that the algorithm is unable to identify the fold bifurcation marking the end of the bistable range; in fact, the estimated branch of unstable solution also exists for $c_3<c_3^*$, with a rapidly increasing amplitude. This phenomenon is due to the slower dynamics (although not critically slow) characterizing the parameter space preceding the fold. Indeed, this effect is exploited in \cite{habib2023predicting, kadar2024model} to predict the fold bifurcation itself.
Nevertheless, this drawback of the algorithm is conservative concerning stability. Therefore, if properly taken into account, it can also be probably accepted in real applications.

\subsection{Larger dimensional systems} \label{sect_multiDoF}

For a single-DoF, two-dimensional system, implementing the ULC estimation method is significantly simpler than for a larger dimensional system.
Since the system is two-dimensional, the so-called primary spectral submanifold \cite{haller2016nonlinear, cenedese2022data} coincides exactly with the full phase space. Accordingly, after any perturbation, the system will still be on the primary spectral submanifold and converge in a spiraling motion toward the equilibrium (unless the perturbation moves the system outside of the ULC).

This is no longer valid in a larger dimensional system, leading to various challenges.
% the ULC typically corresponds to a single dimensional object laying on the boundary of the equilibrium solution's basin of attraction. The basin of attraction is often defined, at least partially, by the attractive manifold of the ULC. By itself, this is not a problem for the method, 
First, a perturbation might bring the system to a region of the phase space far away from the ULC, from which the system might converge toward the equilibrium without approaching the ULC. Consequently, it might carry only minimal information regarding the ULC.
Assuming that any perturbation can be applied to the system, there is no straightforward way to solve this issue. However, dynamical systems tend to converge towards the primary spectral submanifold, which, for a vibrating system, is often a two-dimensional invariant surface. As explained later, this phenomenon organizes the dynamics, making the ULC estimation usually possible.

Another important challenge of two or more DoF systems is that they might undergo modal interaction and beating phenomena, significantly complicating the ULC estimation.
Currently, we cannot provide any solution for this issue, which remains an open topic and limits the method to system having one dominant vibration mode.
This limitation will be addressed in future studies.

Despite these problems, the method is, in general, applicable to systems having dimension larger than two, as illustrated in Sections \ref{sect_delay} and \ref{sect_PP}.
In the following, the main differences that should be implemented with respect to the analysis of a two-dimensional system are presented.

We assuming that a trajectory converges towards the primary spectral submanifold while approaching the equilibrium; therefore, presents a sort of ordered dynamics. The first part of the trajectory is, in general, transient while approaching the spectral submanifold. Accordingly, it should be discarded. The time length of the portion of the trajectory to be discarded cannot be defined a priory, since it strongly depends on the initial conditions.
One way to solve the problem is the following.
The wavelet transformation of one of the system coordinates' time series is computed, providing its frequency content at each time step.
At each time step, the frequency with the largest content corresponds to the mode of interest. Conversely, the frequency with the second largest frequency content (the second local peak of the signal coming from the wavelet transformation at each time instant) corresponds to the second mode, which might disturb the computation.
We assume that the part of the time series to be discarded last until the second largest energy content frequency is above a given threshold. The time series is used for the analysis only once the second local peak is below this threshold.
A simpler approach consists of choosing a predefined time length of the trajectory to be discarded. This can be selected through a visual inspection of the signal and modifying it in a trial-and-error manner.

For the prediction of the ULC, each system coordinate is taken independently. The trajectory is projected on a position-velocity two-dimensional subspace of the phase space. In this plane, Poincar\'e sections are defined and the prediction is carried out as for the single-DoF case.
Since the analysis is performed on a projection of the trajectory intersections are possible, and can compromise the estimation.
If a sufficiently long initial part of the signal is neglected, and the topology of the system is relatively simple, this problem does not subsist.

If a mathematical model of the system is available, either from the system mechanics or obtained from data, instead of using the measured coordinates, it is possible to use modal coordinates. However, since the developed method is explicitly developed to overcome the need for a mathematical model, this case is not considered here.

In the case of a single-DoF, two-dimensional system, only the first three logarithmic decrement/amplitude points are collected and used to compute a quadratic interpolation, whose intersection with the zero axis indicates the estimated ULC amplitude.
For larger dimensional systems, it is convenient to use more points and compute a least square approximation to identify the quadratic curve. In this way, the disturbance due to other modes potentially present in the signal is partially mitigated. The same approach should be used in the case of real systems to mitigate noise. However, the experimental validation of the method is not discussed here, but it is left for future studies.

In the following sections, the method is applied to three inherently different systems, in order to evaluate its potentiality and limitation.
Namely, a mass-on-moving-belt system, which presents non-smoothness; a single-DoF model of turning machining, whose dynamics is described by a delayed differential equation, leading to an infinite dimensional system; and a pitch-and-plunge two-DoF model of an airfoil undergoing aeroelastic flutter.

\section{Non-smooth system: Mass-on-moving-belt}\label{sect_MOB}

\begin{figure*}[t]
\begin{center}
\setlength{\unitlength}{\textwidth}
\begin{subfigure}[b]{0.325\textwidth}
    \includegraphics[trim={0 -8cm 0 0},clip,width=\textwidth]{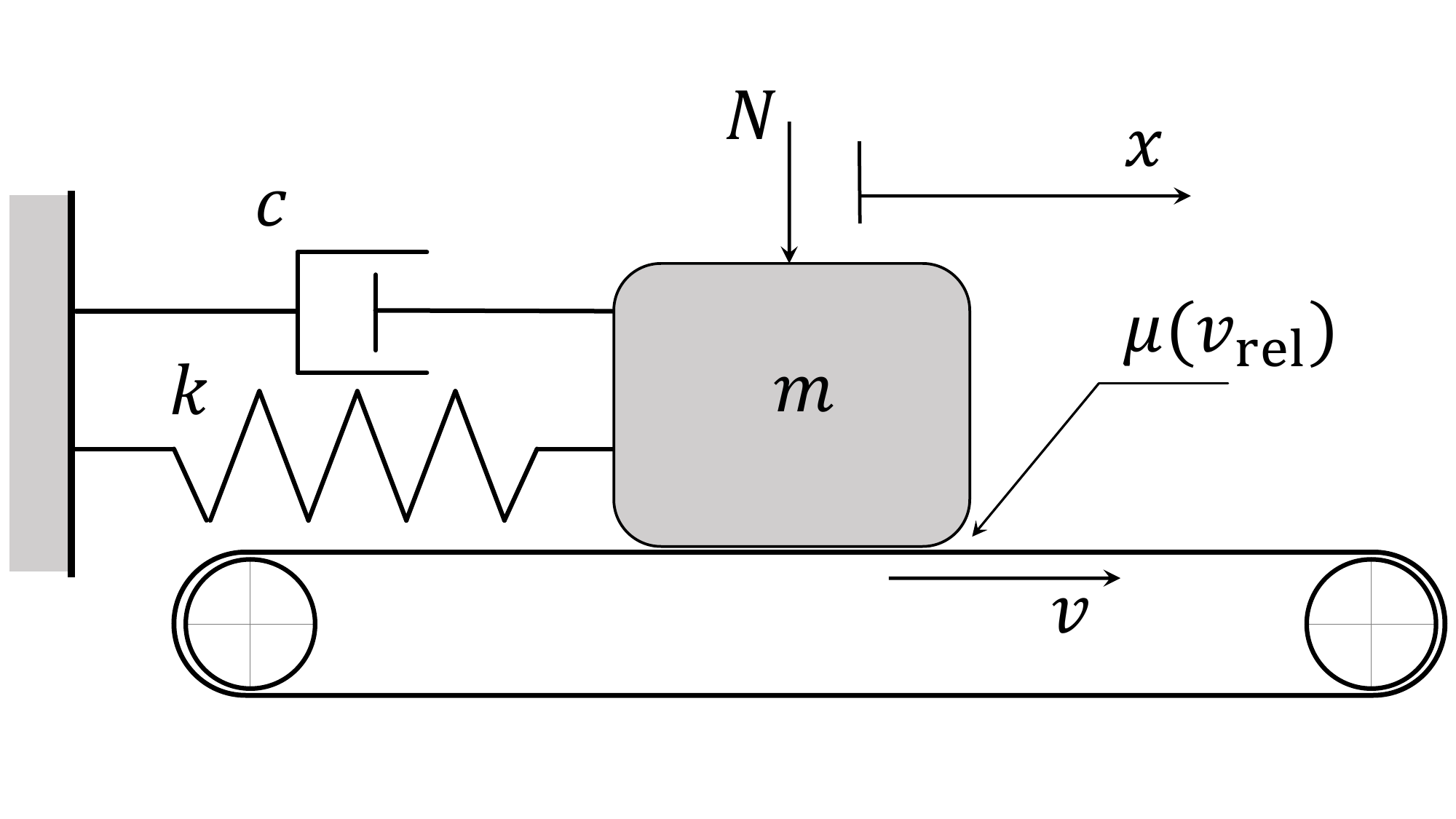}
    \caption{}
    \label{fig_model_MOB}
  \end{subfigure}
\begin{subfigure}[b]{0.325\textwidth}
    \includegraphics[width=\textwidth]{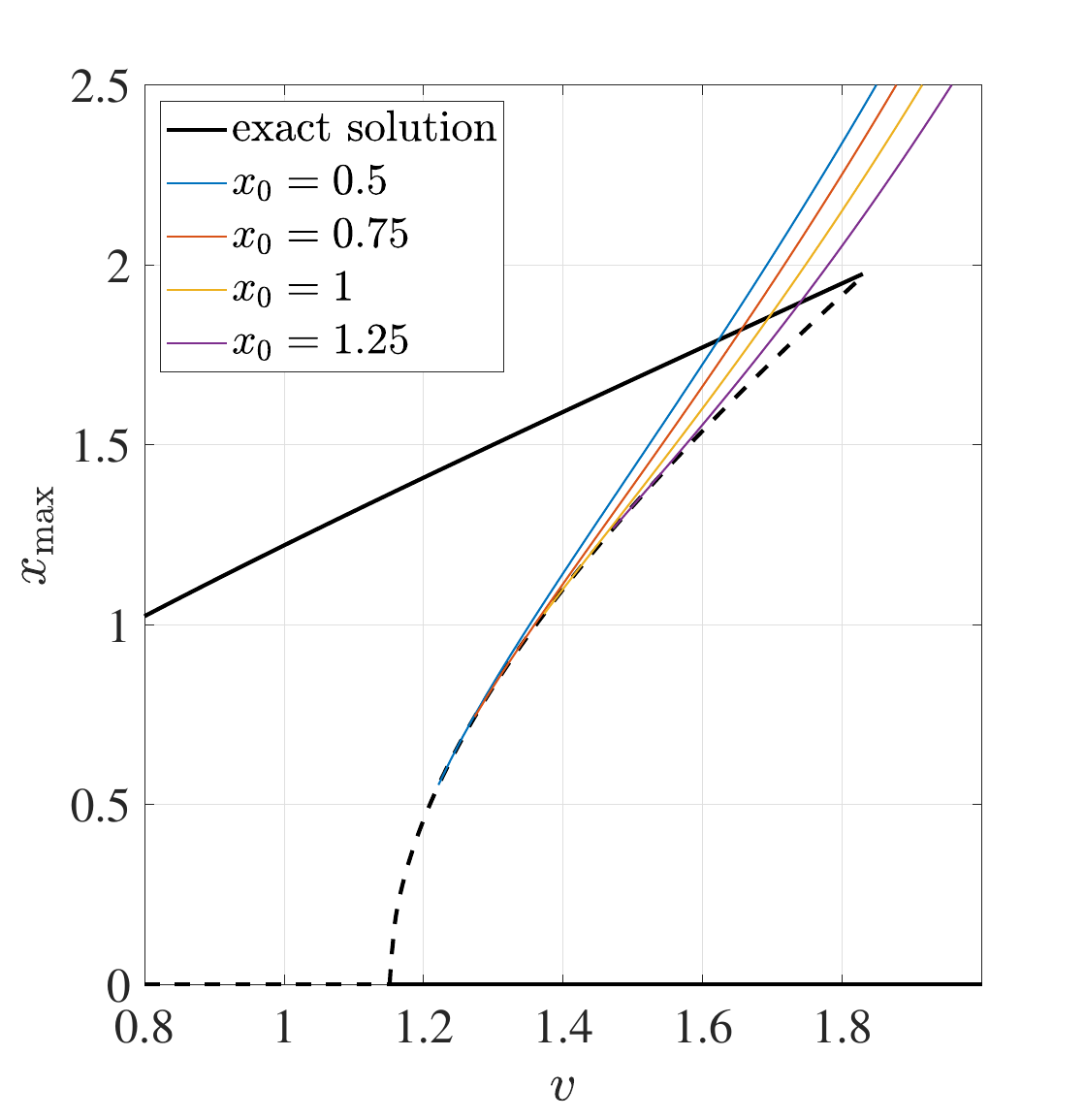}
    \caption{}
    \label{fig_bif_diag_MOB}
  \end{subfigure}
\begin{subfigure}[b]{0.325\textwidth}
    \includegraphics[width=\textwidth]{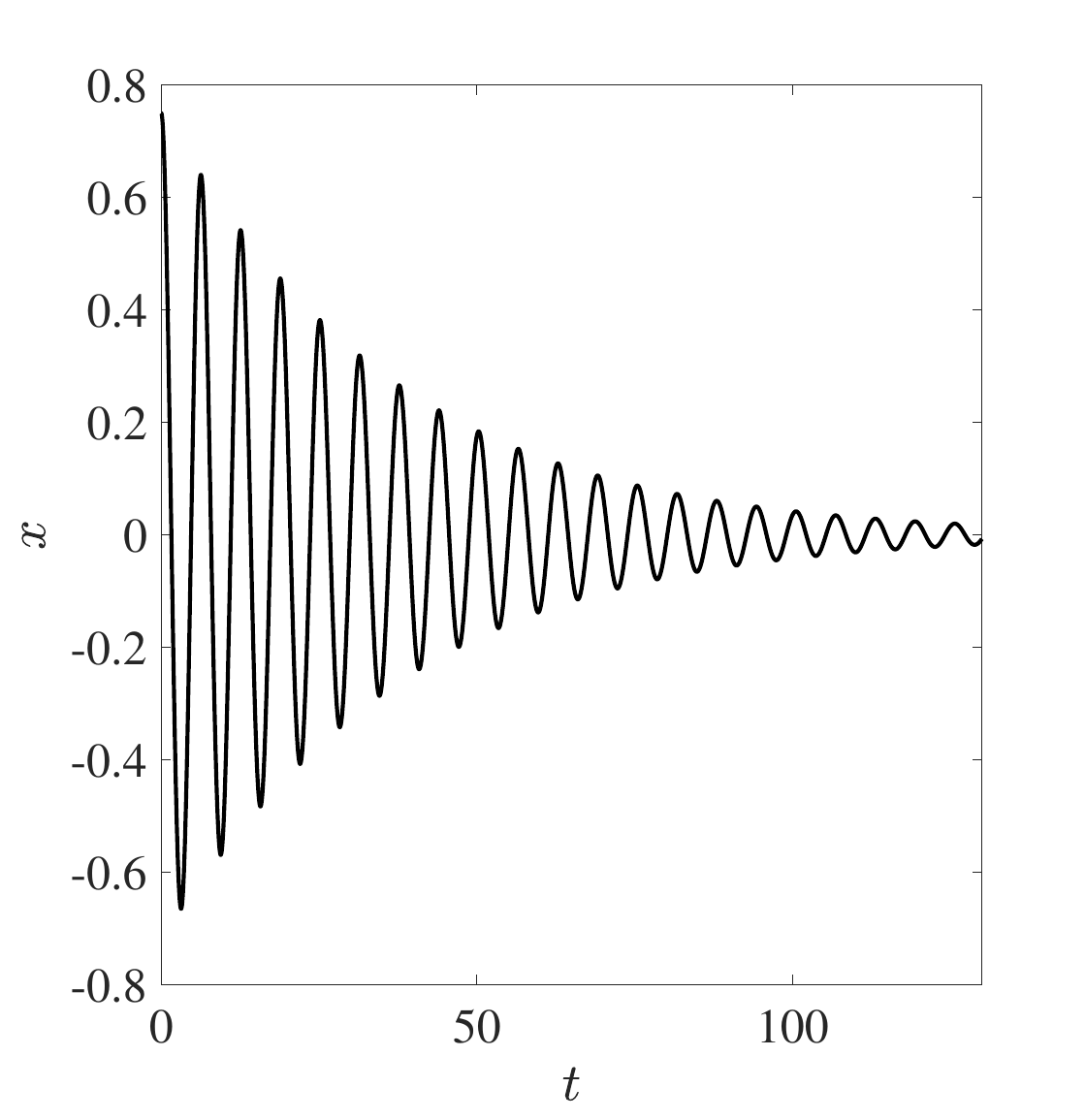}
    \caption{}
    \label{fig_TS_MOB}
  \end{subfigure}
  \begin{subfigure}[b]{0.325\textwidth}
    \includegraphics[width=\textwidth]{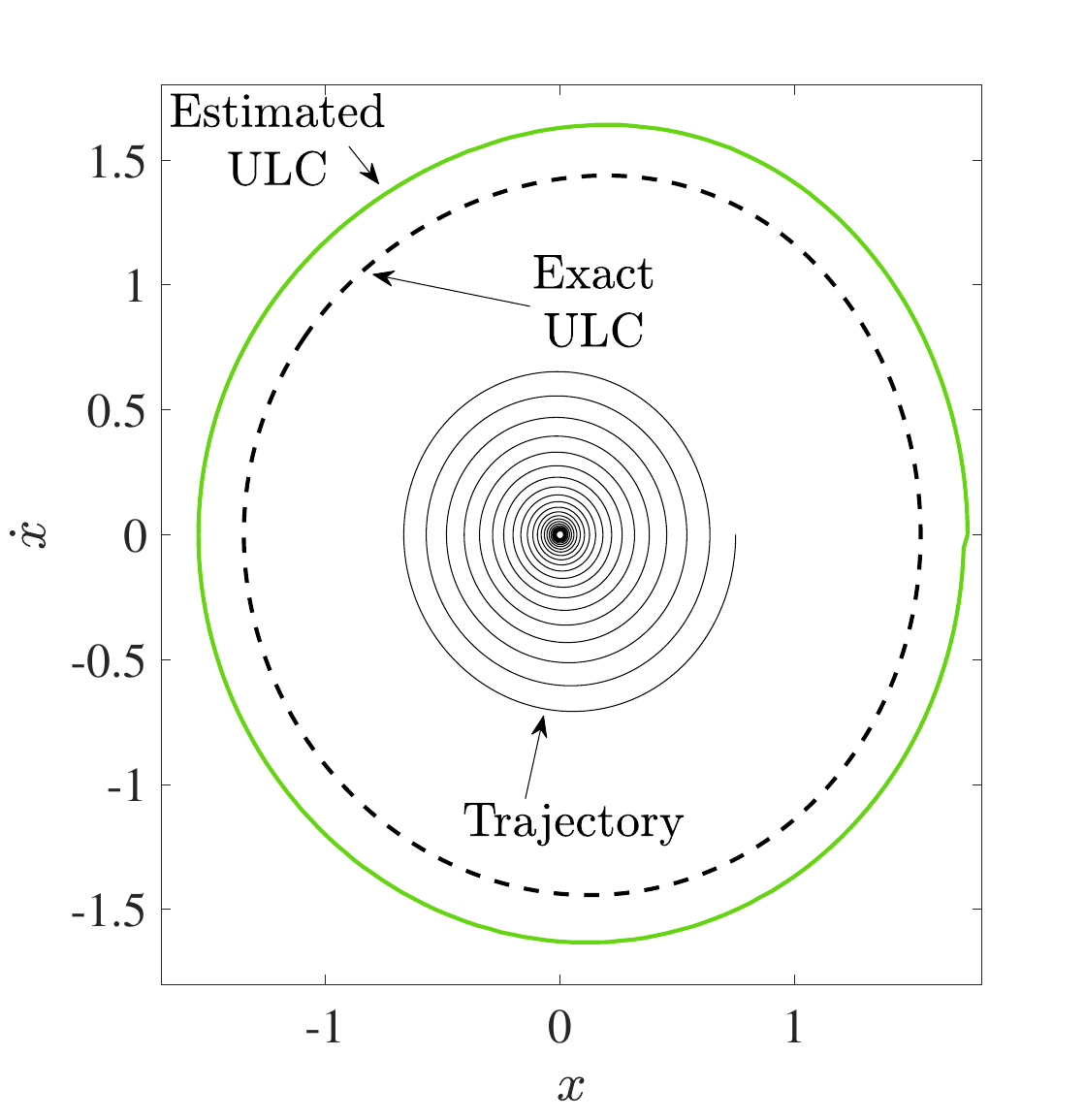}
    \caption{}
    \label{fig_PP_MOB}
  \end{subfigure}
  \begin{subfigure}[b]{0.325\textwidth}
    \includegraphics[width=\textwidth]{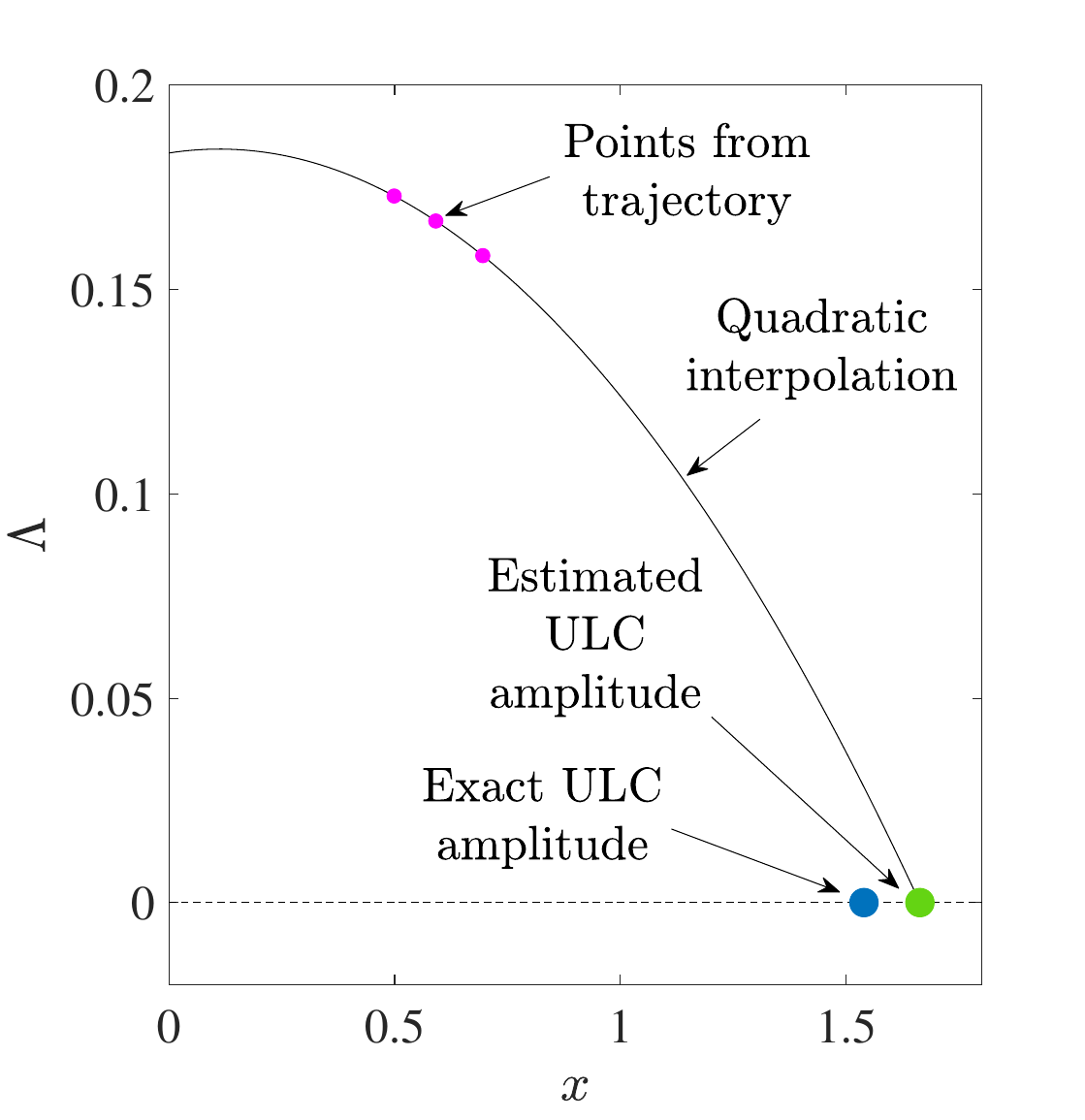}
    \caption{}
    \label{fig_LD_MOB}
  \end{subfigure}
  \begin{subfigure}[b]{0.325\textwidth}
    \includegraphics[width=\textwidth]{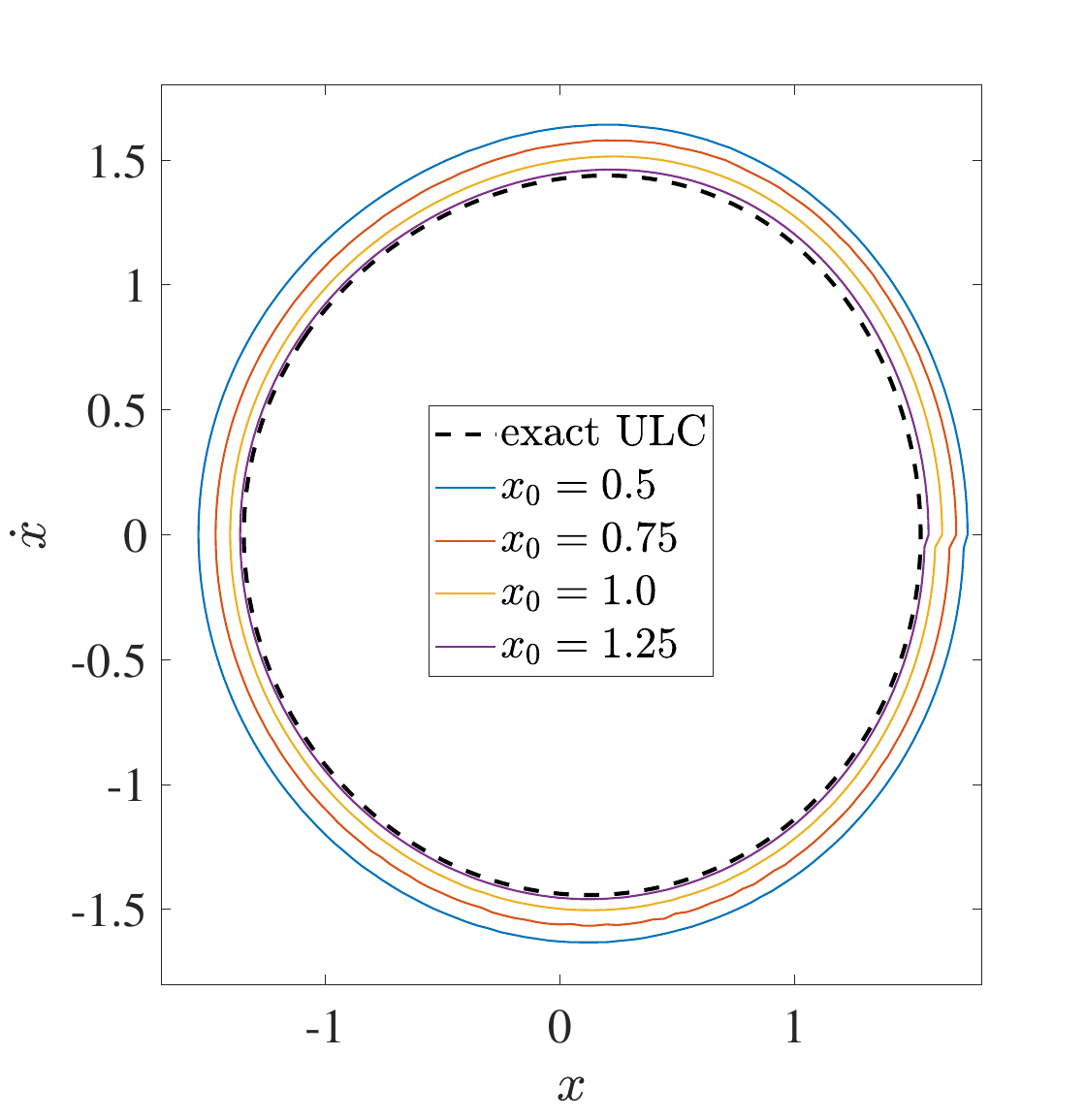}
    \caption{}
    \label{fig_PP_pred_MOB}
  \end{subfigure}
\end{center}
\caption{\label{fig_MOB}Comprehensive results about the mass-on-belt system. (a) Mechanical model. (b) Bifurcation diagram; black curves: exact solution; colored lines: estimated. (c) Time series for the ULC estimation, obtained for $v=1.6$. (d) Phase portrait of the system for $v=1.6$; dashed curve: exact ULC; green curve: estimated ULC. (e) Logarithmic decrement points and ULC estimation. (f) Comparison between exact and estimated ULC for different initial conditions.}
\end{figure*}

We consider the classical mass-on-moving-belt system represented in Figure \ref{fig_model_MOB} \cite{papangelo2017subcritical, hu2020friction}, which is an archetypal model for friction-induced vibrations, typically used for studying violin string \cite{leine1998stick} and brake squeal dynamics \cite{hetzler2007steady}.

The non-dimensional equation of motion of the system is \cite{papangelo2017subcritical} \begin{equation}
\ddot x+2\zeta\dot x+x=F_{\text f},\label{eq_MOB}
\end{equation}
where $\zeta$ is the linear damping ratio and
\begin{equation}
\left\{\begin{array}{ll}
F_{\text f}=\mu\left( v_{\text{rel}}\right)&\quad v_{\text{rel}}\neq0\\
|F_{\text f}|\leq\mu_{\text s}&\quad v_{\text{rel}}=0\,.
\end{array}\right.
\end{equation}
According to Stribeck friction law \cite{jackobson2003stribeck}, the friction coefficient $\mu$ is given by the exponential decaying function \begin{equation}
\mu (v_{\text{rel}}) = \left(\mu_{\text d}+\left(\mu_{\text s}-\mu_{\text d}\right) \text e^{-\frac{|v_{\text{rel}|}}{v_0}}\right) \text{sign} \left(v_{\text{rel}}\right) \,,
\end{equation}
where $v_{\text{rel}} = v - \dot x $ is the relative velocity.
The friction force presents a discontinuity for $v_{\text{rel}}=0$.
For this study, the same parameter values utilized in \cite{hu2020friction} are adopted, i.e., $\mu_{\text s}=1$, $\mu_{\text d}=0.5$, $v_0=0.5$ and $\zeta=0.05$.

As thoroughly explained in \cite{papangelo2017subcritical}, the system's equilibrium loses stability through a subcritical Andronov-Hopf bifurcation for a specific belt velocity $v=v_{\text{cr}}=1.151$.
The equilibrium is stable for $v>v_{\text{cr}}$ and unstable otherwise.
A branch of unstable periodic solutions arises at $v=v_{\text{cr}}$.
This branch then merges with a branch of stable periodic solutions through a saddle-node bifurcation for $v=1.83$. The merging of the two branches is non-smooth, as illustrated in Figure \ref{fig_bif_diag_MOB}.

We apply the ULC estimation technique for this system.
The system is two-dimensional; therefore, the procedure outlined in Sect.~\ref{sect_meth_1DoF} can be directly implemented.
However, the system is non-smooth, which might challenge the estimation algorithm.
We fix $v$ at 1.6, and we consider initial conditions $x(0)=x_0$ and $\dot x(0)\approx0$.
Initially, we set $x_0=0.75$, which is approximately halfway between the ULC and the trivial equilibrium.
The obtained time series is illustrated in Fig.~\ref{fig_TS_MOB}.
Considering the local maximum of the trajectory, we estimate an amplitude of 1.665 for the ULC, while the exact amplitude is 1.539; accordingly, we register an overestimation error of about 8~\% (Figs.~\ref{fig_PP_MOB} and \ref{fig_LD_MOB}).
Repeating the procedure with other initial conditions, according to expectations, the error increases for lower initial conditions (12~\% error for $x_0=0.5$) and decreases for larger initial conditions, closer to the ULC (4~\% error for $x_0=1.0$ and about 1~\% error for $x_0=1.25$), as shown in Fig.~\ref{fig_PP_pred_MOB}.
Overall, considering that the procedure's objective is to provide a rough estimation of the ULC, the result can be considered satisfactory.
We also note that the estimated ULC has the same asymmetric shape as the exact ULC.

Repeating the same procedure for various values of the belt velocity $v$, i.e., the bifurcation parameter, the branch of ULCs can be estimated.
The same perturbation amplitudes are considered, namely $x_0=0.5$, 0.75, 1.0 and 1.25.
The larger the ULC is, the worse the estimation is expected to be because the trajectory is farther from the ULC.
This phenomenon is confirmed by the results obtained, as illustrated in Fig.~\ref{fig_bif_diag_MOB}.
The branch of ULCs is better estimated when its amplitude is small.
Also in this case, the estimation algorithm is unable to identify the saddle-node bifurcation, and the estimated ULC branches are prolonged beyond the saddle-node bifurcation in a region where no ULC exists.
Additionally, no variation in the trend of the ULC branches is visible, unlike in the previous case studied.
This observation confirms the limitation of the method discussed above.

We note that, although this system is non-smooth, the non-smoothness becomes particularly relevant mostly for the stick-slip periodic solutions beyond the ULC. Therefore, it only affects the estimation procedure mar\-gi\-na\-lly. Conversely, a strong non-smooth phenomenon acting between the trajectory and the ULC to be estimated is expected to compromise the method's effectiveness completely.

\section{System with time delay: turning machining}\label{sect_delay}

\begin{figure*}[t]
\begin{center}
\setlength{\unitlength}{\textwidth}
\begin{subfigure}[b]{0.325\textwidth}
    \includegraphics[trim={0 -4cm 0 0},clip,width=\textwidth]{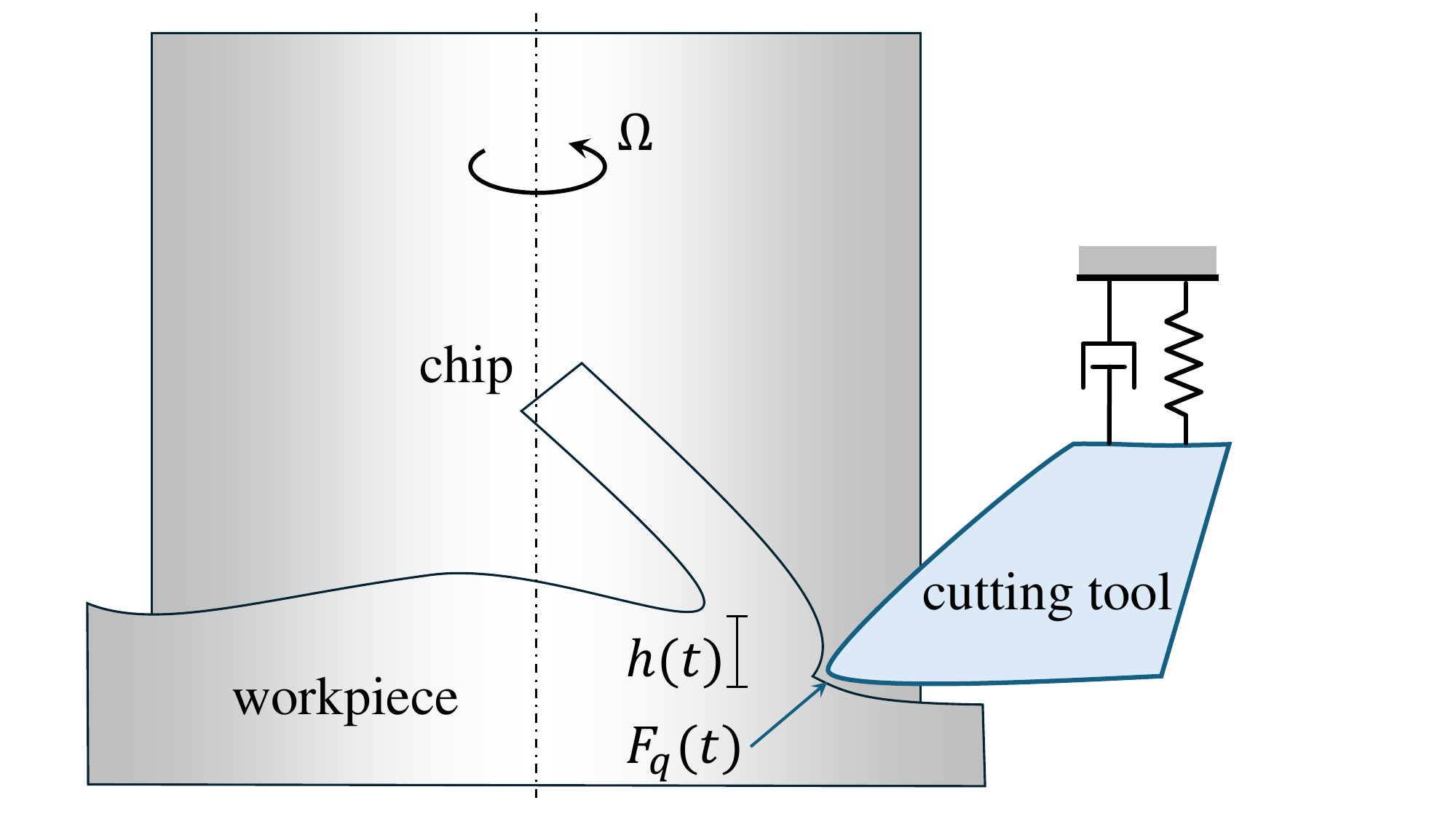}
    \caption{}
    \label{fig_model_CH}
  \end{subfigure}
\begin{subfigure}[b]{0.325\textwidth}
    \includegraphics[width=\textwidth]{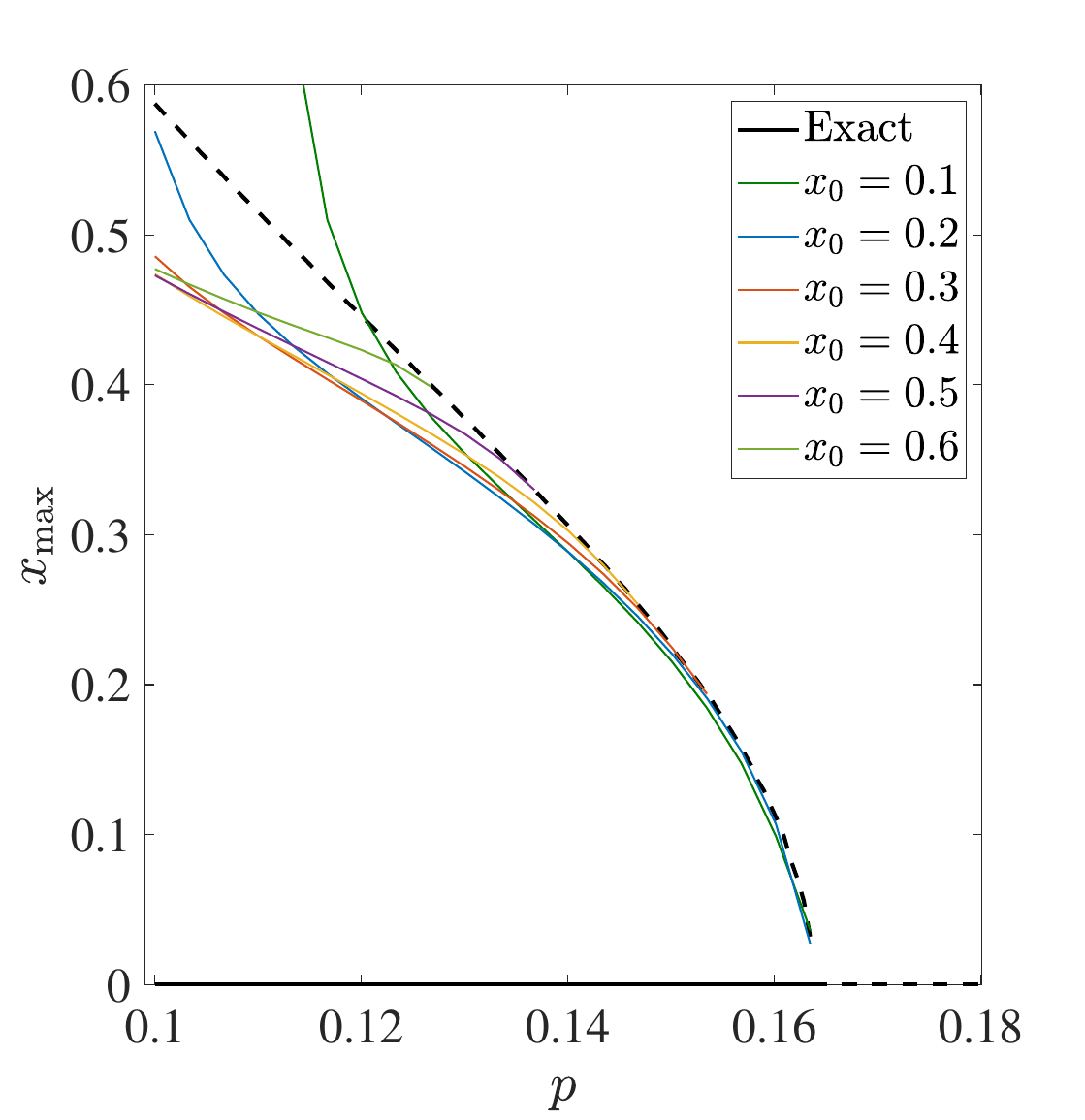}
    \caption{}
    \label{fig_bif_diag_CH}
  \end{subfigure}
\begin{subfigure}[b]{0.325\textwidth}
    \includegraphics[width=\textwidth]{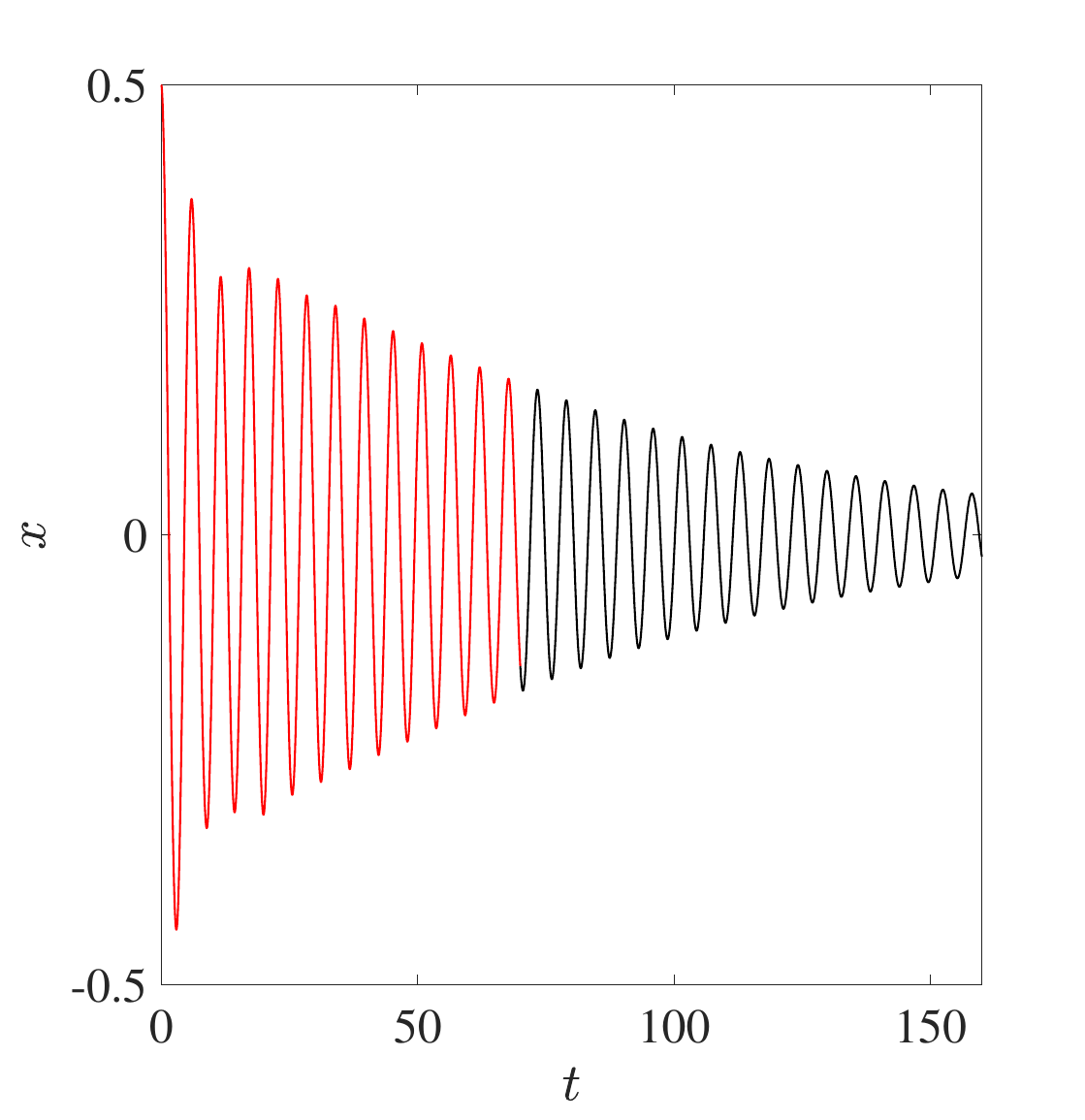}
    \caption{}
    \label{fig_TS_CH}
  \end{subfigure}
  \begin{subfigure}[b]{0.325\textwidth}
    \includegraphics[width=\textwidth]{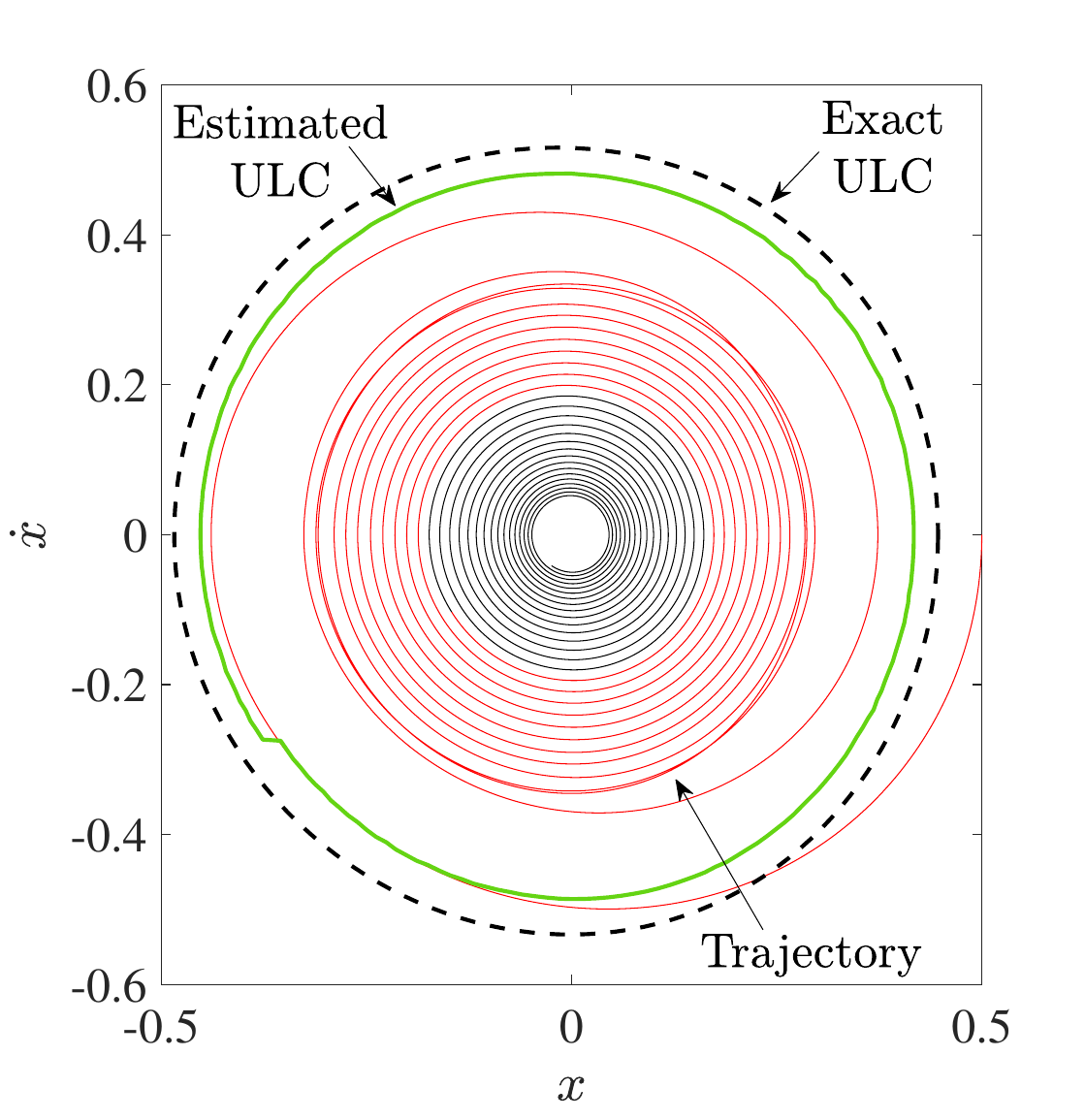}
    \caption{}
    \label{fig_PP_CH}
  \end{subfigure}
  \begin{subfigure}[b]{0.325\textwidth}
    \includegraphics[width=\textwidth]{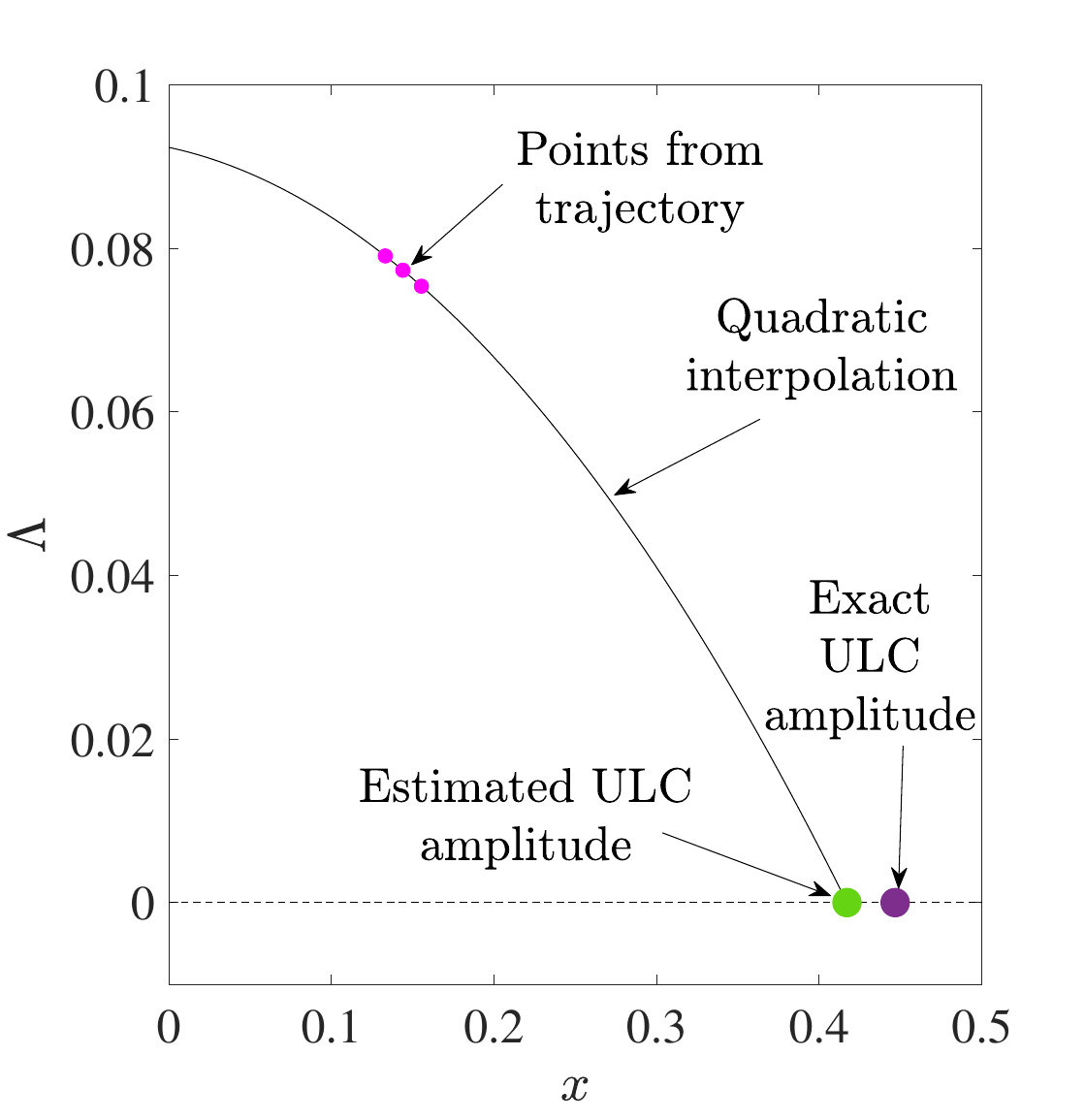}
    \caption{}
    \label{fig_LD_CH}
  \end{subfigure}
  \begin{subfigure}[b]{0.325\textwidth}
    \includegraphics[width=\textwidth]{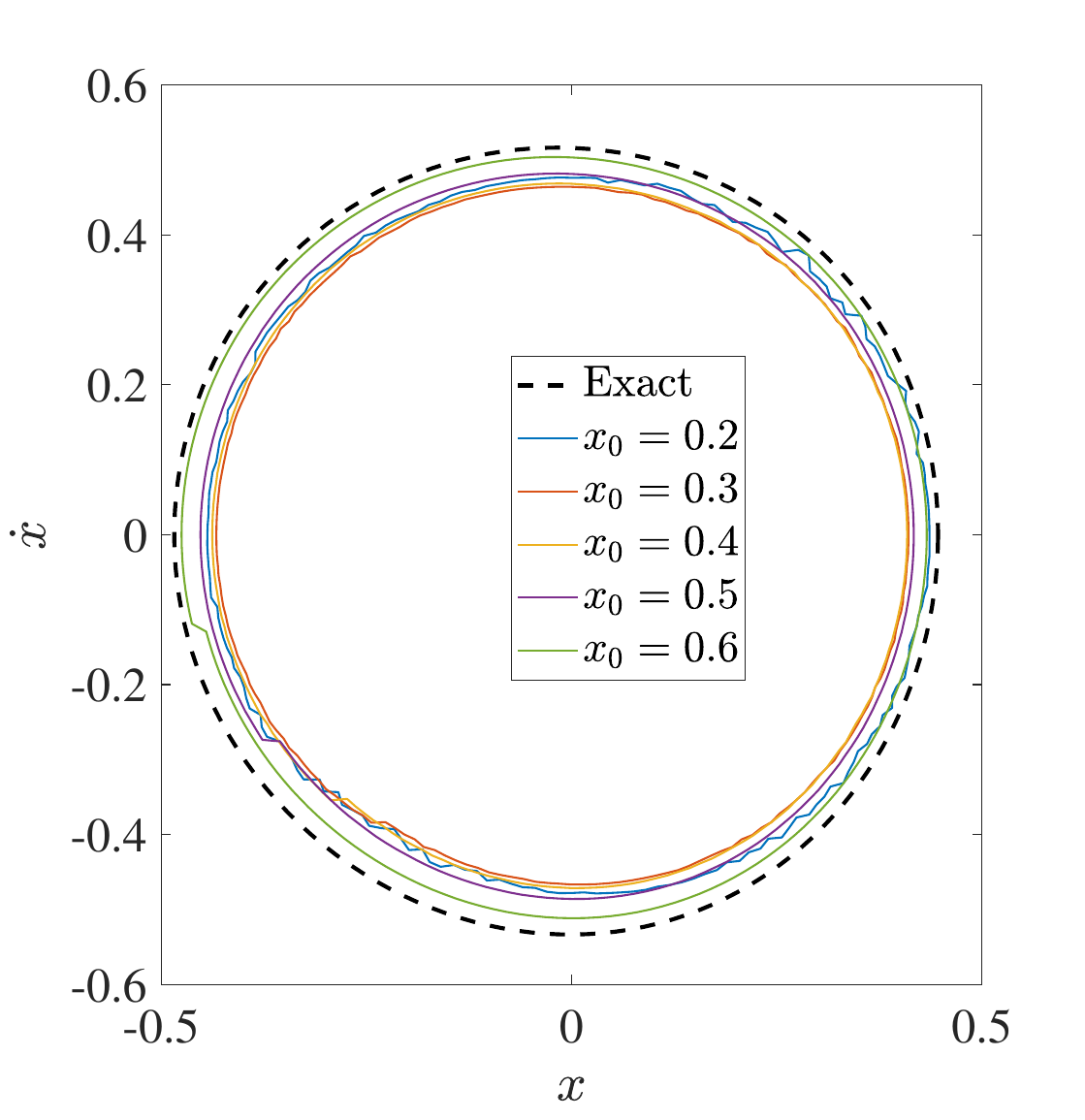}
    \caption{}
    \label{fig_PP_pred_CH}
  \end{subfigure}
\end{center}
\caption{\label{fig_CH}Comprehensive results about the turning machining model. (a) Mechanical model. (b) Bifurcation diagram; black curves: exact solution; colored lines: estimated. (c) Time series for the ULC estimation, obtained for $v=1.6$. (d) Phase portrait of the system for $v=1.6$; dashed curve: exact ULC; green curve: estimated ULC. (e) Logarithmic decrement points and ULC estimation. (f) Comparison between exact and estimated ULC for different initial conditions.}
\end{figure*}

We consider a simplified model of a turning machine, as represented in Fig.~\ref{fig_model_CH}.
The cutting tool is modeled as a single-DoF oscillator, where the mass, damping, and stiffness coefficients represent the modal mass, damping, and stiffness, respectively, which are non-di\-men\-sio\-na\-lized in this study.
The work-piece, assumed stationary, interacts with the cutting tool through the cutting force $F_c$, which depends on the depth of cut.
The time-dependent depth of cut $h(t)$ is given by the difference in the position of the cutting tool between the present position and its position one revolution earlier. This phenomenon introduces a time delay in the differential equation describing the system dynamics, which is, therefore, a delay differential equation.

The time delay in the equation of motion makes the system infinite dimensional, despite having only one-DoF.
Consequently, the system will be studied according to the methodology discussed in Sect. \ref{sect_multiDoF}, as illustrated below.

The depth of cut has nominal value $h_0$, which is also the value it assumes in the case of steady cutting.
Calling $x$ the modal displacement relative to the cutting tool's first mode, normalized with respect to the nominal depth of cut, and normalizing the time with respect to the cutting tool's first natural frequency, the equation of motion is \begin{equation}
\begin{split}
&\ddot x(t)+2\zeta\dot x(t)=p\big(\left(x(t-\tau)-x(t)\right)\\
&\quad+\eta_2\left(x(t-\tau)-x(t)\right)^2+\eta_3\left(x(t-\tau)-x(t)\right)^3\big),
\end{split}
\end{equation}
where $\zeta$ is the damping factor, $\tau$ is the dimensionless spindle period of rotation (and the dimensionless time delay of the system), $p$ is the dimensionless chip width, and $\eta_2$ and $\eta_3$ describe the nonlinear characteristic of the cutting force function.
In particular, \begin{equation}
\begin{split}
&\eta_2=h_0\frac{\rho_2+3\rho_3 h_0}{\rho_1+2\rho_2 h_0+3\rho_3 h_0^2}\\
&\eta_3=h_0^2\frac{\rho_3}{\rho_1+2\rho_2 h_0+3\rho_3 h_0^2},
\end{split}
\end{equation}
where $\rho_1=6109.6$~Nmm$^{-2}$, $\rho_2=-54141.6$~Nmm$^{-3}$ and $\rho_3=203769$~Nmm$^{-4}$ are coefficient identified experimentally in \cite{shi1984theory}. 
Although different models for the cutting force exist and their validity is an open
research topic \cite{dombovari2008estimates}, they all agree on the nonlinear nature of the cutting force with respect to the chip thickness.
Without loss of generality, in this study, the coefficient values $\zeta=0.05$, $\tau=9$, $h_0=0.07$~mm are used.
More details about the mathematical model adopted can be found in \cite{habib2017chatter}.
We note that the so-called fly-over effect \cite{beri2020essential}, which occurs when the cutting tool leaves the work-piece, leading to a non-smoothness in the cutting force value, is neglected.
Considering that the model is used only for demonstration purposes about the method's effectiveness and does not aim at providing new results about turning dynamics, this simplification does not interfere with the scope of the paper.

This system exhibits a classical subcritical An\-dro\-nov-Hopf bifurcation, as illustrated in Fig.~\ref{fig_bif_diag_CH}.
In this case, no stable periodic solution is present according to the mathematical model. In reality, if too large perturbations are given to the system, periodic, quasiperiodic, or chaotic motions are generated, involving the fly-over effect \cite{beri2020essential}.
The branch of ULCs extends until its amplitude does not exceed the chip thickness, which marks the maximal range of validity of the mathematical model.

For the generation of time series, the utilized initial function is \begin{equation}
\begin{array}{ll}
x(t)=0,\,\dot x(t)=0&\text{ for }t\in\left(-\tau,0\right)\\
x(t)=x_0,\,\dot x(t)=0&\text{ for }t=0.
\end{array}
\end{equation}
Initially, we fix $p=0.12$. 
For this $p$ value, trajectories converge to the equilibrium for $x_0<0.6986$; otherwise, they diverge.

The ULC is initially estimated for $x_0=0.5$. Since the system has a large (infinite) dimension, the first 70 time units of the time series are neglected in order to give time to the system to converge to the vicinity of the primary spectral submanifold.
The length of the time series to be neglected was chosen through a trial-and-error approach. Slightly shorter or longer lengths of the neglected portion of the time series do not significantly affect the estimation.

The obtained time series is illustrated in Fig.~\ref{fig_TS_CH}, where the red curve represents the neglected part of the time series, while only the black part is used for the estimation.
The trajectory is represented in the phase space in Fig.~\ref{fig_PP_CH}.
The estimated ULC, represented in green in Fig.~\ref{fig_PP_CH}, is relatively close to the exact one depicted by the black dashed curve. The relative error for $\dot x=0$ is approximately 6.6~\%.
We note that, although the red part of the trajectory is not so far from the ULC and cannot be considered a small perturbation, the black part is relatively far from the ULC.

Figure \ref{fig_LD_CH} shows the logarithmic decrement-amplitude points picked from the trajectory at $\dot x=0$. The difference between the exact and the estimated ULC amplitude from the quadratic interpolation can be recognized in the figure.

The procedure is further validated utilizing different initial conditions, down to $x_0=0.2$. The relative error obtained for the different cases (computed for $\dot x=0$) is 2.3~\% for $x_0=0.2$, 8.1~\% for $x_0=0.3$, 8.4~\% for $x_0=0.4$, and 3.1~\% for $x_0=0.6$. 

Finally, Fig.~\ref{fig_bif_diag_CH} compares the estimated branches of ULC, obtained with different amounts of perturbations, with the exact one. All branches provide an acceptable estimation, except for $x_0=0.1$, which estimates very large amplitudes for $p<0.115$.

\section{Multi-DoF system: aeroelastic flutter} \label{sect_PP}

\begin{figure}[h]
\begin{center}
\setlength{\unitlength}{\textwidth}
    \includegraphics[trim={2cm 0 0 0},clip,width=0.4\textwidth]{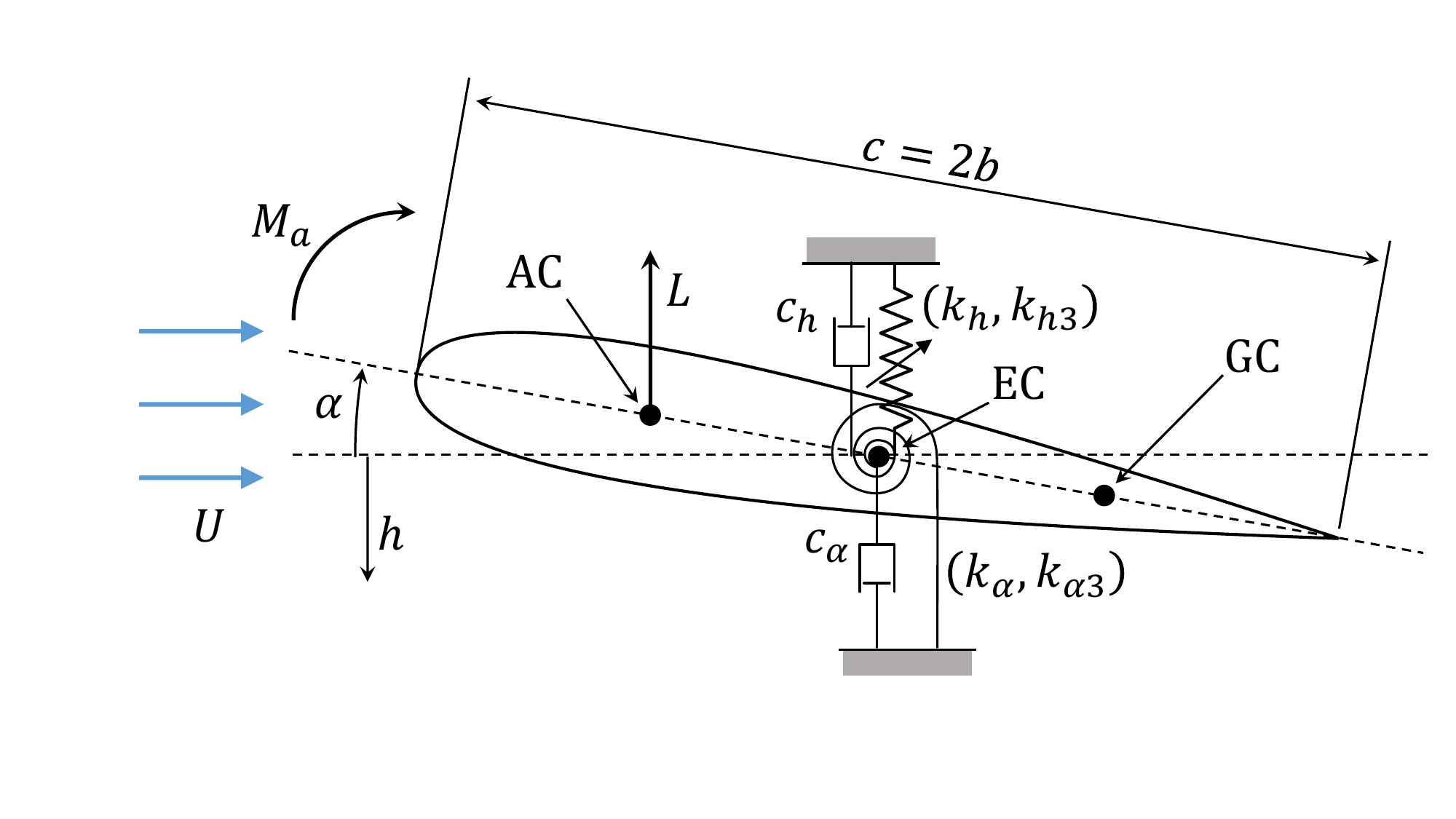}
\end{center}
\caption{\label{fig_model_PP}Mechanical model of the pitch-and-plunge airfoil.}
\end{figure}

The method is applied on a pitch-and-plunge wing profile exposed to airflow and undergoing flutter oscillations (Fig.~\ref{fig_model_PP}); the system is similar to the one studied in \cite{malher2017flutter}.
The pitch-and-plunge wing model was implemented in various previous studies \cite{dowell2014modern, lee2007suppressing1, lee2007suppressing2} with only slight variations from the one considered here regarding the system nonlinearities.
Differently from the systems studied in the previous sections, this model has two DoFs, which makes it four-dimensional.
Accordingly, it should be analyzed through the method described in Sect.~\ref{sect_multiDoF}.

Non-dimensional equations of motion governing the dynamics of the system are \begin{equation}
\mathbf M\ddot{\mathbf x}+\mathbf C\dot{\mathbf x}+\mathbf{Kx}+\mathbf b\left(\mathbf x\right)=\mathbf 0,\label{eq_EOM_PnP}
\end{equation}
where
\begin{equation}
\begin{split}
&\mathbf x=\left[\begin{array}{c}
y\\\alpha
\end{array}\right],\,\mathbf K=\left[\begin{array}{ccc}
\Omega^2&\beta u^2\\
0&r_\alpha^2-\nu u^2
\end{array}\right],\,\mathbf C=\left[\begin{array}{ccc}
\zeta_h+\beta u &0\\
-\nu u&\zeta_\alpha
\end{array}\right],\\
&\mathbf M=\left[\begin{array}{ccc}
1&x_\alpha\\x_\alpha &r_\alpha^2\end{array}\right],\,\mathbf b=\left[\begin{array}{c}
0\\
\xi_{\alpha3}\alpha^3+\xi_{\alpha5}\alpha^5
\end{array}\right],
\end{split}
\end{equation}
$\alpha$ marks the pitch rotation, and $y$ indicates the heave displacement, non-di\-men\-sion\-al\-ized in relation to the semichord of the airfoil, while $u$ is the non-dimensional flow velocity.
More information about the physical meaning of all the other parameters can be found in \cite{malher2017flutter}; a similar nomenclature is utilized here for facilitating the comparison.
The adopted parameter values are $x_\alpha=0.2$, $r_\alpha=0.5$, $\beta=0.2$, $\nu=0.08$, $\Omega=0.5$, $\zeta_\alpha=0.01$, $\zeta_h=0.01$, $\xi_{\alpha3}=-1$ and $\xi_{\alpha5}=20$.

\begin{figure*}[t]
\begin{center}
\setlength{\unitlength}{\textwidth}
\begin{subfigure}[b]{0.325\textwidth}
    \includegraphics[width=\textwidth]{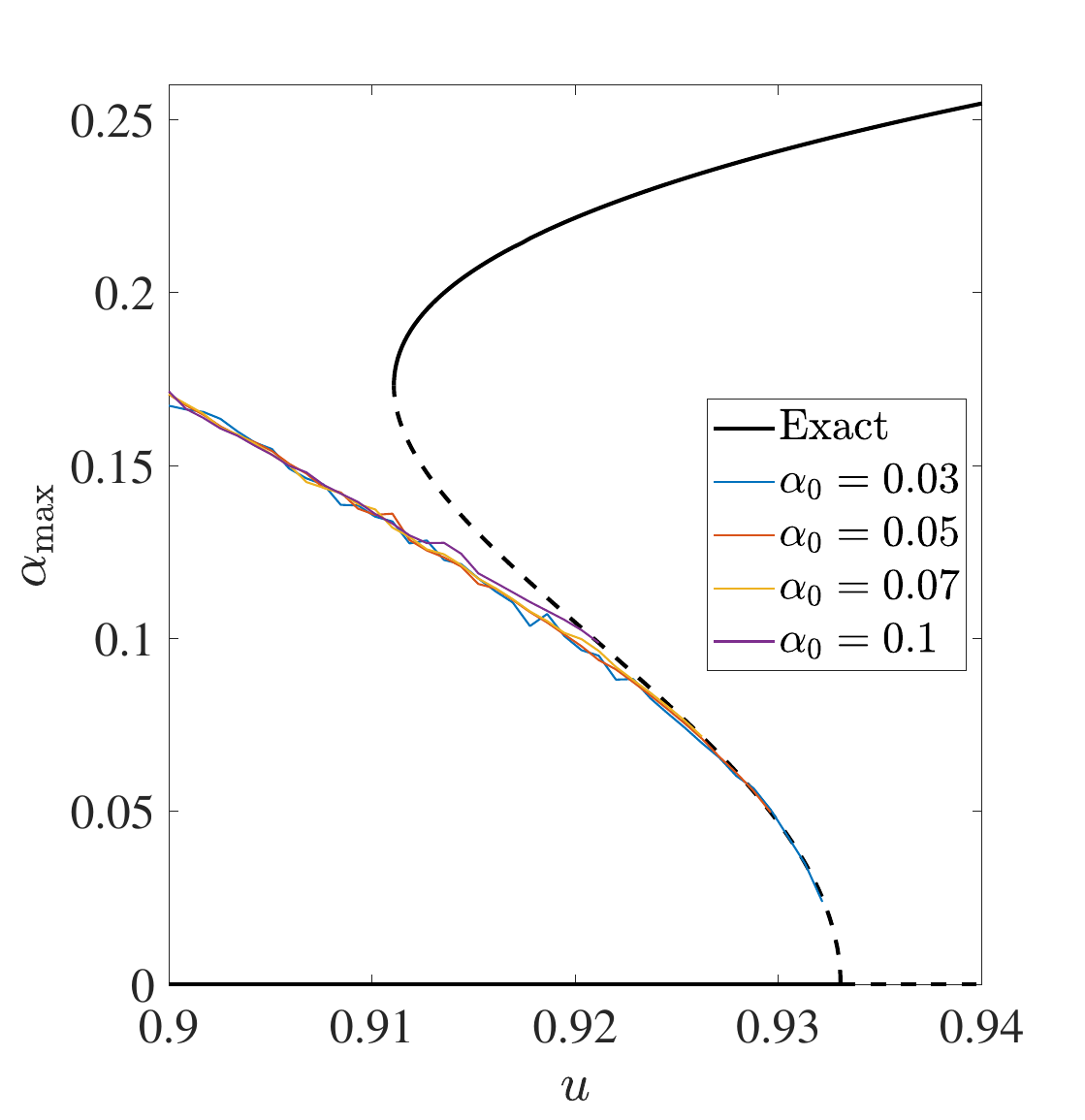}
    \caption{}
    \label{fig_bif_diag_PP}
  \end{subfigure}
\begin{subfigure}[b]{0.325\textwidth}
    \includegraphics[width=\textwidth]{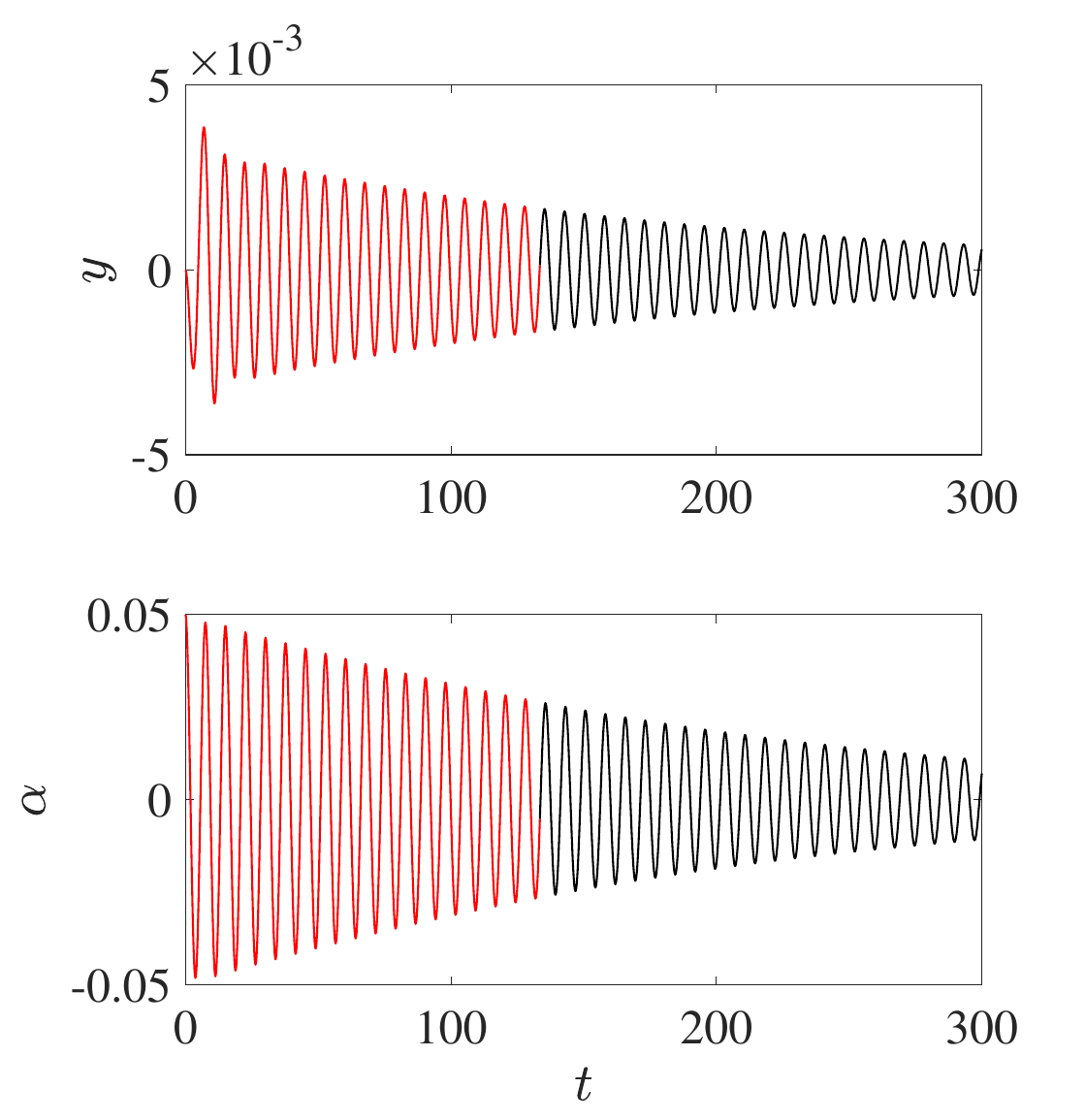}
    \caption{}
    \label{fig_TS_PP}
  \end{subfigure}
  \begin{subfigure}[b]{0.325\textwidth}
    \includegraphics[width=\textwidth]{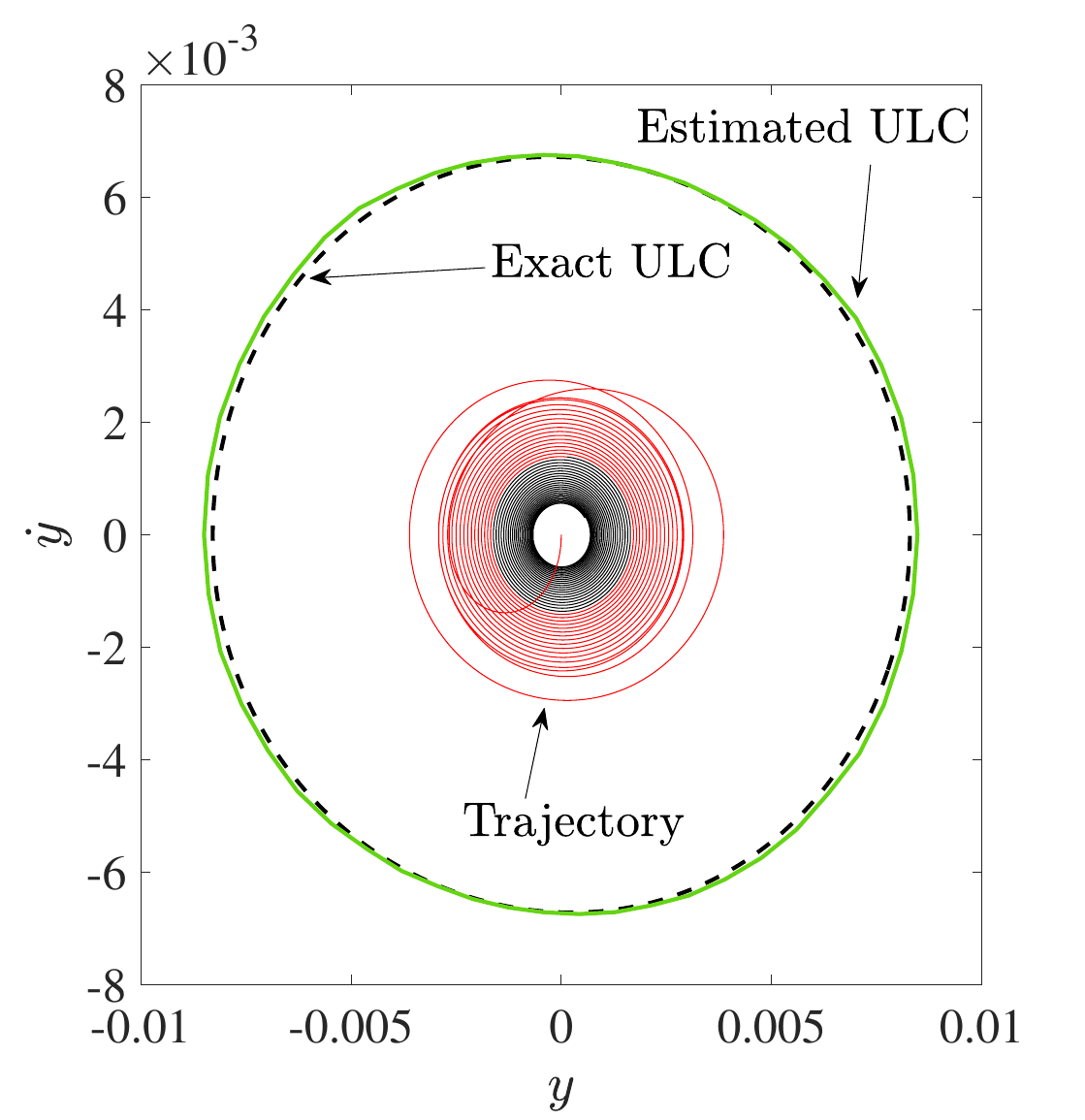}
    \caption{}
    \label{fig_PP_PP_y}
  \end{subfigure}
  \begin{subfigure}[b]{0.325\textwidth}
    \includegraphics[width=\textwidth]{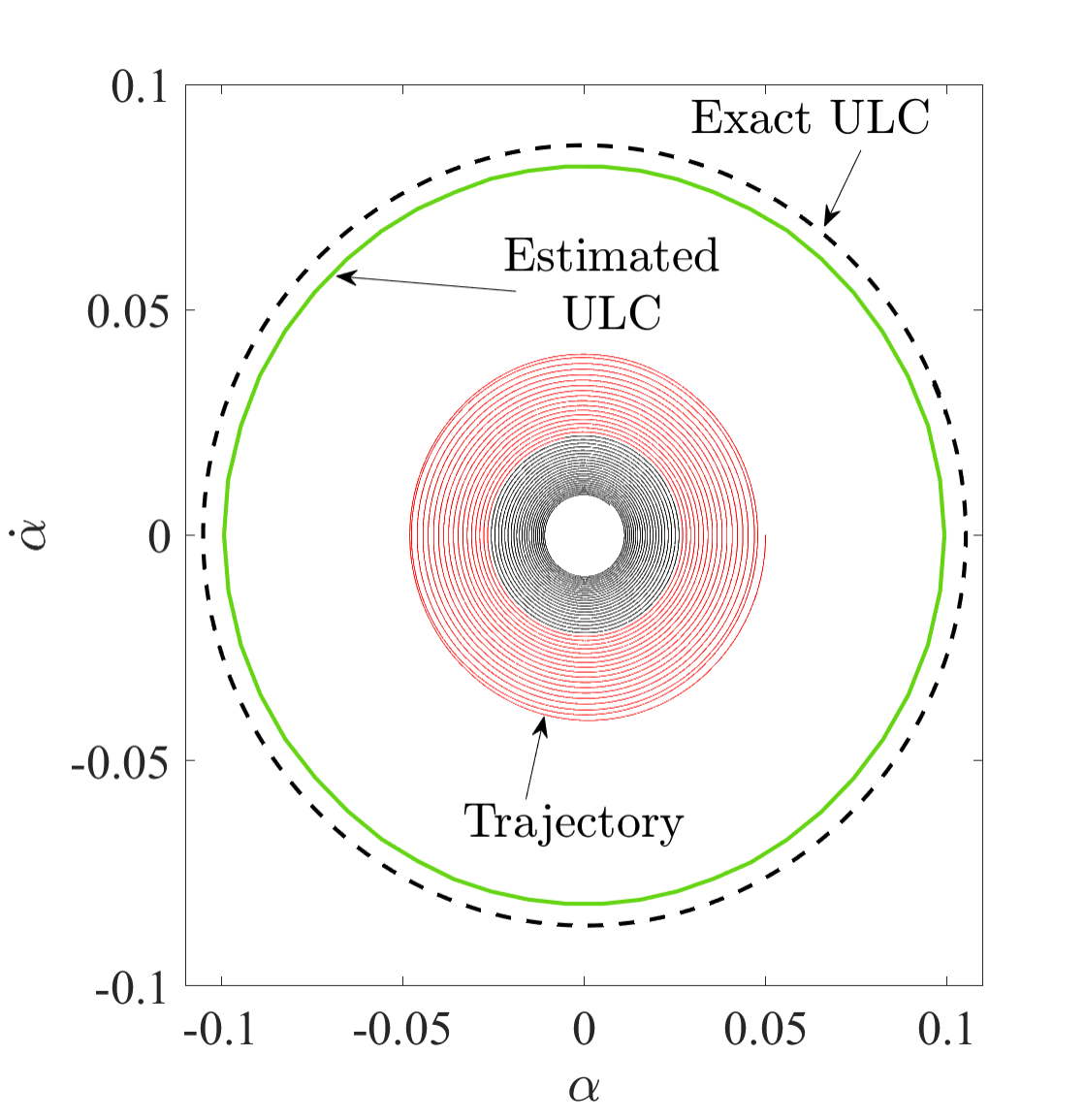}
    \caption{}
    \label{fig_PP_PP}
  \end{subfigure}
  \begin{subfigure}[b]{0.325\textwidth}
    \includegraphics[width=\textwidth]{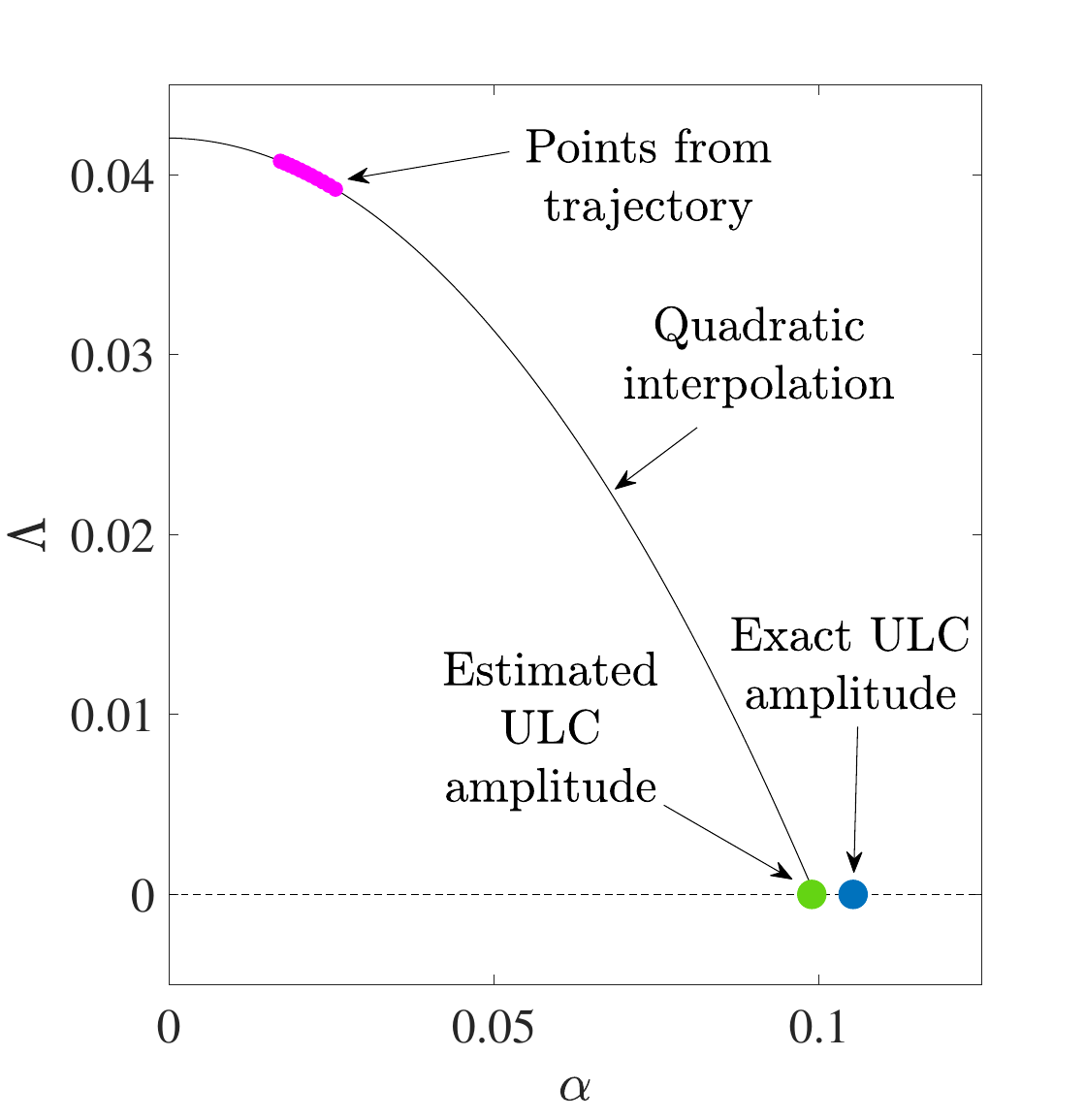}
    \caption{}
    \label{fig_LD_PP}
  \end{subfigure}
  \begin{subfigure}[b]{0.325\textwidth}
    \includegraphics[width=\textwidth]{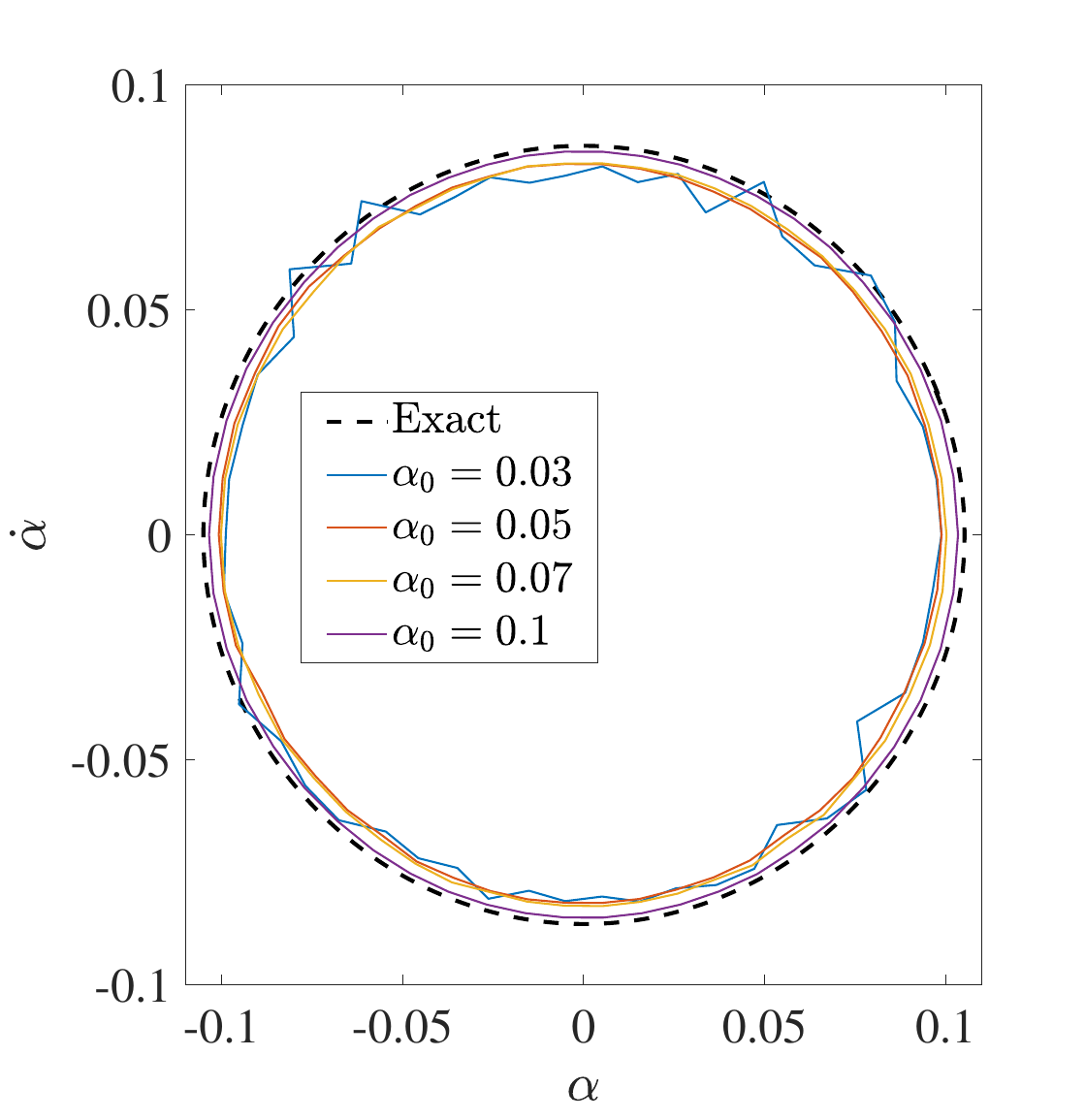}
    \caption{}
    \label{fig_PP_pred_PP}
  \end{subfigure}
\end{center}
\caption{\label{fig_PP}Comprehensive results about the pitch-and-plunge wing profile system. (a) Bifurcation diagram. (b) Time series for the ULC estimation, obtained for $u=0.92$; red curve: discarded trajectory; black curve: used trajectory (c,d) Phase portrait of the system for $u=0.92$ ($y$ and $\alpha$ coordinates, respectively); red curve: discarded trajectory; black curve: used trajectory; dashed curve: exact ULC; green curve: estimated ULC. (e) Logarithmic decrement point and ULC estimation. (f) Comparison between exact and estimated ULC for different initial conditions.}
\end{figure*}

For the considered parameter values, the bifurcation diagram for variations of the flow velocity $u$ is shown in Figure \ref{fig_bif_diag_PP}.
The equilibrium position loses stability through a subcritical Andronov-Hopf bifurcation; the emerging branch of unstable periodic solutions turns back at a fold for $u=u^*$ and becomes stable.
The Andronov-Hopf bifurcation occurs at $u=0.933$, while the fold is at $u=u^*=0.911$.

Next, we try to estimate the ULC for $u=0.92$.
We simulate a trajectory of the system with initial conditions $y(0)=0$, $\alpha(0)=\alpha_0$, $\dot y(0)=0$, and $\dot \alpha(0)=0$.
For a first trial, we impose $\alpha_0=0.05$. The obtained time series are illustrated in Fig.~\ref{fig_TS_PP}, while the corresponding trajectory in the phase space is shown in Figs.~\ref{fig_PP_PP_y} and \ref{fig_PP_PP} in two different projections.
The trajectory encompasses both the red and black curves in the figures. However, the red curve is the discarded part, and only the black one is utilized for the computation.
To select the portion of the trajectory to be discarded, the wavelet transformation of the pitch displacement is computed. Then, for each time instant, the second largest local maximum of the frequency is computed, which is strictly related to the energy content of the system's second mode. When this value was below 0.00001, we considered that the signal contains energy in only one mode. The threshold value was chosen arbitrarily through a trial-and-error procedure.
We note that increasing the threshold to 0.0001 provided similar results to those shown below.

Applying the procedure outlined in Sect.~\ref{sect_multiDoF}, we first consider the plunge motion $y$ and its derivative $\dot y$. Results of the estimation are illustrated in Fig.~\ref{fig_PP_PP_y}, where the black dashed line indicates the exact ULC, while the green line the estimated one. The matching between the two is excellent (error 2.1~\%).
Fig.~\ref{fig_PP_PP} presents the results concerning the pitch DoF $\alpha$. In this case, results are slightly less accurate but sufficiently good for the purposes of the proposed methodology (error 6~\%).

Figure \ref{fig_LD_PP} illustrates the estimated logarithmic decrement with respect to the peaks of the displacement in pitch. The figure compares the estimated ULC amplitude (green dot) and the exact one (blue dot). We note that in this case, 11 logarithmic decrement-amplitude points are utilized to compensate for disturbance generated by the second vibration mode of the system.
The quadratic curve was found through a least square approximation.

Figure \ref{fig_PP_pred_PP} shows the results obtained with different initial conditions (in the $\alpha$-$\dot\alpha$ space). We note that too small initial conditions ($\alpha_0=0.03$) generate a fuzzy curve due to numerical integration inaccuracies and approximations in obtaining the correct intersection of the trajectory with the Poincar\'e maps.
However, all considered cases produced satisfactory results.
The obtained relative errors are 6~\% for $\alpha_0=0.03$, 4.7~\% for $\alpha_0=0.07$, and 1.7~\% for $\alpha_0=0.1$.

Finally, the colored curves in Fig.~\ref{fig_bif_diag_PP} indicate the estimated bifurcation diagram for the four considered initial conditions of the signals.
In all cases, the estimation is satisfactory until the fold bifurcation.
Differences between the four cases are minimal because, although initial conditions are different, the threshold for eliminating contribution from other modes was the same for all.
The algorithm fails to recognize that before the fold, no ULC exists, highlighting a limitation already discussed in the previous cases.

\section{Discussion and conclusions}\label{sect_conclusions}

This study introduced a new method for estimating unstable limit cycles (ULCs) surrounding stable equilibrium solutions.
The method leverages the critical slowing down of the system dynamics in the vicinity of a steady-state solution, either stable or unstable.
Recognizing that the logarithmic decrement is an effective way to measure the system's slowing down, the method enables one to estimate a ULC from a trajectory initiated by a relatively small perturbation from the equilibrium.

The method was tested on various systems, each presenting different challenges.
Apart from a relatively simple case of a resonator with nonlinear damping, we investigated the mass-on-moving-belt system, which pre\-sents non-smoothness, a turning machining model, whose equation of motion includes time delay, and a pitch-and-plunge airfoil undergoing flutter, which has two-DoF (differently from the other systems which have only one-DoF).
The method proved to be effective in all cases.

The main advantages of the method are: \begin{itemize}
\item It does not require any sort of mathematical model of the system. This is a significant advantage with respect to other data-driven methods, which usually have predictive properties only through a mathematical model, either physics-based or black box. Bypassing the model generation and its interpretation, some potential sources of error are avoided and computational cost is significantly reduced.
\item For implementing the method to real systems, there is no need for actuators or feedback controllers. Other experimental methods for identifying unstable solutions require a feedback control, such as the control-based optimization method \cite{renson2019numerical, raze2024experimental}, or at least actuators. The method developed in this study might be effective even if it exploits only natural perturbations to the system experienced during normal operation.
\item The method does not require that the system's dynamics is in close vicinity of the unstable limit cycle, but it estimates its position from a certain distance in the phase space. However, the closer it is to the ULC, the more accurate the estimation.
\item The method requires only a single time series for the estimation, which makes it computationally much cheaper than the large majority of data-driven methods and easily implementable in experimental systems.
\end{itemize}

Regarding the limitations of the system, we note:
\begin{itemize}
\item The method exploits a quadratic interpolation for estimating the amplitude of the ULC. However, there is no guarantee that the logarithmic decrement follows a quadratic trend. In fact, this is not generally the case.
This might lead to significant errors. However, this approximation proved to be a simple and effective way for the estimation, providing a balance between simplicity and precision of computation.
\item The method cannot deal with a non-smooth system if the non-smoothness is between the provided trajectory and the ULC to be estimated. However, as illustrated in Sect.~\ref{sect_MOB}, in general, non-smoothness does not necessarily prevent the method from working.
\item In its present form, the method cannot deal with systems presenting strong modal interaction.
\item Multi-DoF systems pose a significant challenge to the methods. Although, as illustrated in Sect.~\ref{sect_PP}, the method can generally work with those kinds of systems, it is required to neglect the first part of the signal. The estimation is rather sensitive to the neglected signal portion; accordingly, special care should be paid in those cases.
\item The method extrapolates the information encompassed in a trajectory. As such, it is prone to provide wrong results if the system has exotic dynamical behavior. Accordingly, users should be aware that the estimated ULC might be even significantly wrong.
\end{itemize}

Future developments of this work include an experimental validation of the method, which is required to prove that the method can be used in real-life applications.
Additionally, the method will be extended to find ULCs in topologically different phase portraits, i.e., where the ULC is not necessarily surrounding a stable equilibrium.
Finally, a strategy to extract useful information also from the portion of signal discarded in multi-DoF systems will be sought.

%%%%%%%%%%%%%%%%%%%%%%%%%%%%%%%%%%%%%%%%%%%%%%%%%%%%%%%%%%%%%%%%%%%%%%%%%%%%%%%%%%%%%%%%%%%%%%%%%%%%%%%%%%%%%%%%%%%%%%%%%

\begin{acknowledgements}
The author thanks G\'abor St\'ep\'an for extensive discussions.
\end{acknowledgements}

\section*{Declarations}

\section*{Funding}
\noindent 
The research reported in this paper has been supported by Project no.
TKP-6-6/PALY-2021 provided by the Ministry of Culture and Innovation of
Hungary from the National Research, Development and Innovation Fund,
financed under the TKP2021-NVA funding scheme and by the National Research, Development and Innovation Office (Grant no. NKFI-134496).

\section*{Availability of data and material} \noindent Not applicable.

\section*{Code availability}
\noindent The data and the code utilized for generating the presented results are available from the author upon request.

\section*{Conflict of interest}
The author declares that he has no conflict of interest.

\bibliographystyle{spmpsci}      % mathematics and physical sciences
\bibliography{references}   % name your BibTeX data base

\end{document}